\documentclass[12pt]{article}
\usepackage{graphicx}
\usepackage{amssymb}
\usepackage{amsmath}
\usepackage{color,cite}
\usepackage{tikz,xcolor,dsfont,scalefnt}
\usepackage{multirow,slashed}
\usetikzlibrary{calc,decorations.markings,intersections,shapes.arrows,decorations.markings,snakes}
\usepackage{pgf-pie}
\usepackage{hyperref}

\usepackage{tkz-euclide}
\usetkzobj{all}
  \usepackage{feynmp}

\allowdisplaybreaks

\newcommand{\TeV}{\text{TeV}}
\newcommand{\GeV}{\text{GeV}}
\newcommand{\MeV}{\text{MeV}}
\newcommand{\fm}{\text{fm}}
\newcommand{\mb}{\text{mb}}
\newcommand{\pb}{\text{pb}}

\newcommand{\cm}{\text{cm}}
\newcommand{\barn}{\text{b}}
\newcommand{\cL}{\mathcal{L}}
\newcommand{\met}{\slashed{E_T}}

\newcommand{\ptmiss}{\slashed{\vec{p}_T}}

\newcommand{\LQCD}{\Lambda_\text{QCD}}
\newcommand{\be}{\begin{equation}}
\newcommand{\ee}{\end{equation}}

\newcommand{\msbar}{\overline{\text{MS}}}

\usepackage{hyperref}
\hypersetup{
    colorlinks,
    citecolor=blue,
    filecolor=black,
    linkcolor=blue,
    urlcolor=blue
}

\usepackage{ifpdf}
\ifpdf
\fi

\begin{document} 
\begin{fmffile}{tasifeyn}
\unitlength = 1mm

\thispagestyle{empty}

 \vspace{5mm}

\begingroup\centering
{\Large\bfseries\mathversion{bold}
TASI Lectures on
  Collider Physics
}%

\vspace{7mm}
 
 \begingroup\scshape\large
 Matthew~D.~Schwartz\\
 \endgroup
 \vspace{2mm}
 \begingroup\small
 Department of Physics\\
 Harvard University\\ 
 \endgroup
 \vspace{10mm}

\textbf{Abstract}\vspace{5mm}\par
\begin{minipage}{14.7cm}
These lectures provide an introduction to the physics of particle colliders.
 Topics covered include a quantitative examination of the design and operational parameters of Large Hadron Collider, kinematics
 and observables at colliders, such as rapidity and transverse mass, and properties of distributions, such as Jacobian peaks. In addition,
 the lectures provide a practical introduction to the decay modes and signatures of important composite and elementary and  particles in the Standard Model, from pions to the Higgs boson. The aim of these lectures is provide a bridge between theoretical and experimental particle physics. Whenever possible, results are derived using intuitive arguments and estimates rather than precision calculations. 
\end{minipage}\par
\endgroup

\newpage
\tableofcontents
\newpage

\section{Introduction}
\label{sec:intro}

These lectures provide an introduction to the physics of colliders, particularly focused on the Large Hadron Collider. They are based on summer school lectures given at the Theoretical Advanced Study Institute (TASI) in Boulder, Colorado and at the Galileo Galilei  Institute in Florence, Italy. The target audience is graduate students who have already had some exposure to quantum field theory and particle physics. While the subject of these lectures, collider physics, has much overlap with the perturbative and non-perturbative physics of quantum chromodynamics, those more technical topics are not included (see~\cite{Ellis:1991qj,Schwartz:2013pla,Skands:2011pf} for more information).

Typically, collider physics is not covered in most quantum field theory sequences. Collider physics is a practical subject: how does one extract information from the data at a collider? What are the kinds of things that can be observed, and what do we learn from their observation? To answer these questions,
I begin in Section~\ref{sec:lhc} with an introduction to colliders, particularly the Large Hadron Collider (LHC). A basic familiarity with the units and design specifications of the LHC is essential for even the most basic understanding of contemporary collider physics. I then proceed to discuss kinematics in Section~\ref{sec:kin} and some of the most important observables at colliders, like invariant mass distributions and transverse mass in Sections~\ref{sec:obs} and~\ref{sec:dist}. Section~\ref{sec:jets} is a  very brief introduction to jets; even though jet physics is one of my favorite collider physics topics, there are already many good reviews of the subject.

 The final part of these lectures in Section~\ref{sec:SM} is about the particles in the Standard Model (SM). 
I wrote this section because I have found that while most graduate students in high energy physics can name the particles in the SM, hardly any of them would recognize one of these particles if they saw it. To really understand particle physics, is important to get to know the actual particles. For this reason,
Section~\ref{sec:SM} comprises a large fraction of these notes: I go one-by-one through the SM particles discussing their properties and signatures.

\section{The Large Hadron Collider \label{sec:lhc}}
Our first task is to answer one of the most essential questions about the LHC: why is it designed the way it is? To get a handle on this question, we can ask what are roughly  the requirements we would need to be able to discover a Higgs boson? We will approach this question using dimensional analysis and some basic particle physics.

The base unit for scattering cross sections at colliders is the barn (b): $1 \barn = 10^{-28}~\text{m}^2$.
The name of this unit of area comes from a joke by Enrico Fermi, that neutrons in nuclear reactors hit
Uranium targets as easily as hitting the broad side of a barn. This joke provides a starting point for dimensional analysis:  the cross section for n-U${}^{235}$ scattering is $\sim 1~\barn$. 
What is the cross section for proton-proton scattering at the LHC? Well,
if the volume of a nucleus scales like the atomic number $Z$, then the radius scales like $r\sim Z^{1/3}$ and so the area like $Z^{2/3}$.
Hence one expects proton-proton scattering to have an inclusive cross section of $235^{-2/3}~\barn = 0.03~\barn = 30~\mb$. 

To convert between length and energy, the formula is
\be
 \frac{1}{\GeV^2} \hbar^2 c^2 = 3.894 \times 10^{-32} \text{m}^2 
\ee
This is a little hard to remember. I prefer
\be
200~ \MeV =  \frac{1}{\text{fm}}
\ee
where fm = femtometer = $10^{-15}$ m. 
This says that the strong interaction scale of QCD, $\LQCD\sim 200$ MeV is the same as the ``radius'' of the proton, $r_p \sim 1$ fm. What does the radius of the proton mean? It means that the
proton scattering cross section should be $\sigma \sim \pi r_p^2 =3~\fm^2 = 3 \times 10^{-30} {\text m}^2 = 0.03~\barn$. So this is consistent with the estimate from
scaling the n-U${}^{235}$ cross section.

So proton-proton scattering happens at the tens-of-millibarns level. There is some energy dependence to this total cross section, but it is fairly weak (logarithmic).
Actually, the total scattering cross section is not even precisely defined theoretically. One can define the inelastic cross section from
the probability for the protons to break apart. However, the elastic cross section, where $p p \to p p$, has an infrared divergence in the forward scattering region:
for protons that glance off each other infinitesimally, you can't tell if the protons have scattered at all. 
In addition, it's possible to have some hard interaction among the partons in the protons but still have exactly two protons in the final state (this is called diffractive
scattering). A trick to get a handle on the  cross section is to use the optical theorem. 
This theorem relates the inclusive inelastic-plus-diffractive scattering total cross section to the imaginary part of the amplitude for forward scattering.
Unfortunately, one can neither measure an exactly forward scattering cross section (because the beam is in the way) or measure an imaginary part.
What is done in practice is to place a detector at very small angle $\theta$ to the beam line (at the LHC one of these detectors is called {\sc totem}) 
to measure $pp\to pp$ scattering. Then the limit $\theta\to 0$ is taken and the interference between the QED and QCD components of the amplitude
is used to find the imaginary part. I'm not convinced that this procedure is on entirely sound theoretical footing, but it seems to give
reasonable answers.

In any case, a number $\sigma \sim 30~\mb$ for the total $pp$ cross section is enough to get us started. Next, we can ask how this compares to the cross section
for some process of interest. For example, what is the rate for $W$ boson production? Well, the typical scale for weak interactions is Fermi's constant
\be
\sigma(pp\to W) \sim G_F\sim \frac{g^2}{m_W^2} \sim \frac{1}{(100~\GeV)^2} = 10^{-6}~\mb  = 1~\text{nb}
\ee
Thus we expect roughly 1 in a million proton collisions will produce a $W$ boson. What about the Higgs boson? Well, as you hopefully know, the dominant 
mechanism for Higgs production is from gluon fusion through a top loop. Thus we expect the Higgs cross section to be down by roughly a loop factor
of $\frac{1}{16\pi^2} \sim 10^{-3}$ from weak interaction cross sections that  proceed at tree level (like $W$ production) . So we estimate
\be
\sigma(pp \to h) \sim  
\begin{gathered}
\resizebox{20mm}{!}{
     \fmfframe(0,0)(0,0){
\begin{fmfgraph*}(40,30)
\fmfstraight
	\fmfleft{L1,L2}
	\fmfright{R}
	\fmf{gluon}{L1,v1}
	\fmf{gluon}{L2,v2}
	\fmf{fermion,tension=0.5}{v1,v2,v3,v1}
	\fmf{phantom,tension=0,label=$t$}{v2,v3}
	\fmf{dashes,label=$h$,l.s=left}{v3,R}
\end{fmfgraph*}
}}
\end{gathered}
 \approx 10^{-3} \times \sigma(pp \to W) \sim  10^{-12}~\barn = 1~\pb
 \label{ggh}
\ee
This means we need to collide 1 billion protons to produce a Higgs. 
At 13 TeV, the total cross section for $pp\to h$ production is closer to 40 pb, but this estimate is not bad.

What kind of luminosity does the LHC need? Well, say we want to see 100 Higgs bosons in a year for a discovery. Let's say we look at
$h \to \gamma \gamma$ decay mode. This mode is clean, but has a $10^{-3}$ branching ratio (see Section~\ref{sec:higgs}). Accounting also for experimental efficiencies,
at say the $\epsilon \sim 10^{-2}$ level, we need
\be
\frac{10^9~\text{collisions}}{\text{1 Higgs}} \times \frac{10^3~\text{Higgs}}{ h\to \gamma\gamma} \times \frac{1}{10^{-2}~\text{eff.}}
 \times \frac{100~h\to \gamma\gamma }{\text{year}}\times\frac{\text{year}} {10^7~ \text{s}}
= 10^{9} \frac{\text{collisions}}{s}= 1~\text{GHz}
\ee
So we need to collide 1 billion protons per second to see 100 Higgs events in a year. How is this done?

\begin{figure}[t]
\begin{center}
\begin{tikzpicture}
\node at (0,0) {\includegraphics[width=0.4\columnwidth]{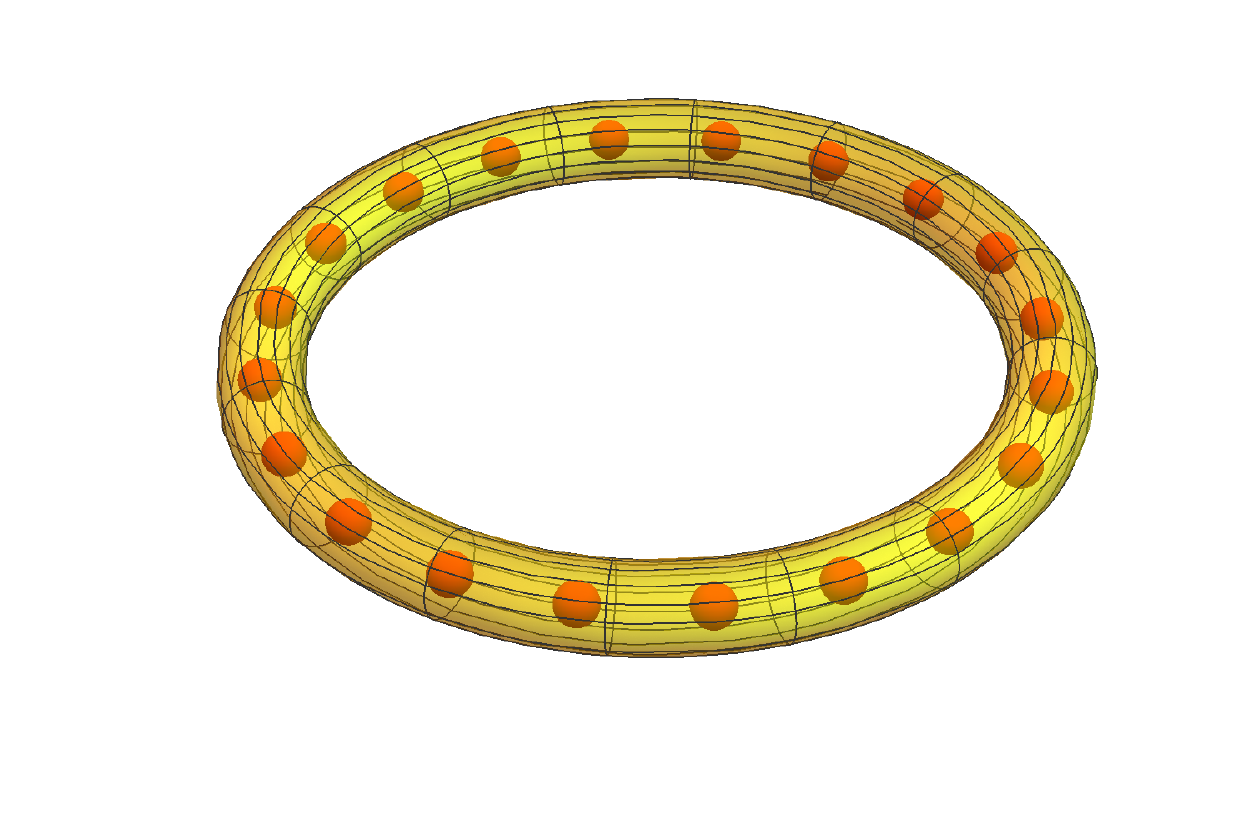}};
\draw [->,line width=2] (-4,0)  to [out=-50, in = 210]  (-2.5,0);
\node [above] at (-5,0)  {$10^{11}$ protons/bunch};
\node [above] at (0,0)  {2800 bunches};
\draw [decorate,decoration={brace,amplitude=4},xshift=-4,yshift=0, line width=2] (3,0.5) -- (3,0) node [black,midway,xshift=80] {bunch spacing = 25 ns};
  \end{tikzpicture}
  \caption{
  The LHC collides protons in bunches. There around 2800 bunches with $10^{11}$ protons per bunch and the bunches collide every 25 ns. 
  }
  \label{fig:bunch} 
\end{center}
\end{figure}

At the LHC the protons are grouped into bunches (see Fig.~\ref{fig:bunch}). The bunches move around the ring at nearly the speed of light,
separated by around 25 ns ($\sim$ 8 meters) from each other. The 25 ns spacing means the bunches collide at 40 MHz. So to get to the GHz rate we need
around 25 collisions per bunch crossing. At the LHC this is achieved by squeezing the bunches to a spot size of around $10$ microns across at the crossing point.
With $10^{11}$ protons per bunch, the number of collisions per bunch crossing is then
\be
N_{\text{events}} =\left(10^{11} \frac{\text{protons}}{\text{bunch}}\right)^2\frac{\sigma_{pp}= 10\,\mb}{\sigma_{\text{beam}}=(10 ~\mu \text{m})^2}
= 100 \frac{\text{collisions}}{\text{bunch crossing}}
\ee
This gives a 4 GHz total collision rate. 

The collision rate at the LHC is called the luminosity. We talk about either integrated or  instantaneous luminosity.
Integrated luminosity are quantities like $25$ fb${}^{-1}$ indicating, for example, that 25 events of a process with a 1 fb cross section should have been produced.
Instantaneous luminosity is what you integrate over time to get the integrated luminosity. 
The instantaneous luminosity of the
LHC is currently $\cL= 10 \frac{\text{Hz}}{\text{nb}} = 10^{34} \frac{1}{\cm^{2}\cdot \text{s}}$. Both nb/Hz and $\cm^{-2}~ \text{s}^{-1}$ are common units
for instantaneous luminosity.  Multiplying by the $pp$ cross section gives $\cL \times 10~ \mb = 10^8$ Hz = 0.1 GHz. This is in the ballpark of what is needed
for Higgs physics. To find exotic beyond-the-standard model physics, the instantaneous luminosity may have to increase by an order of magnitude.

So the LHC is producing around 1 billion events per second. As you probably know GHz is also the speed of current computer processors. When a CPU runs at 1 GHz, 
it can do around 1 billion operations per second. An event from an LHC collision takes up about 1 MB of storage space. Thus it is clearly not possible to write
all 1 billion events to disk every second. Instead, current electronics can record about 200 MB/s to disk. Thus the $10^9$ events must be reduced to about 100 events
to be recorded and analyzed. This reduction is done through the triggering system.

\begin{table}
\begin{center}
\begin{tabular}{|c|c|}
\hline
Trigger & Rate \\
\hline
1 isolated electron, $p_T > 25$ GeV  & \multirow{ 2}{*}{40 Hz}  \\
2 isolated electrons, both $p_T > 15$ GeV & \\
\hline
1 photon, $p_T > 60$ GeV &\multirow{ 2}{*}{40 Hz}  \\
2 photons, $p_T > 20$ GeV & \\
\hline
1 muon, $p_T > 20$ GeV &\multirow{ 2}{*}{40 Hz}  \\
2 muons, $p_T > 10$ GeV & \\
\hline
1 jet, $p_T > 400$ GeV &\multirow{ 3}{*}{25 Hz}  \\
3 jets, $p_T > 165$ GeV & \\
4 jets, $p_T > 110$ GeV &\\
\hline
1 jet, $p_T > 165$ GeV  and $\slashed{E_T} > 70$ GeV& 20 Hz\\
\hline
1 tauon, $p_T > 35$ GeV  and $\slashed{E_T} > 45$ GeV& 5 Hz\\
\hline
2 muons + displaced vertex b-tag& 10 Hz\\
\hline
prescales (e.g.1\% of 1 jet, $p_T > 200$ GeV) & 5 Hz\\
\hline
\end{tabular}
\caption{Example trigger table used by ATLAS or CMS (triggers are frequently updated and changed).
\label{tab:trigg}}
\end{center}
\end{table}

An example trigger table is shown in Table~\ref{tab:trigg}. Each line shows a criterion which, if satisfied, leads  to the event being recorded to disk.
The total rate in the table is 200 Hz, which is around the current capacity of the electronics. 
The triggering is done in stages. The first stage is a low-level hardware trigger that looks for things like hard electrons or photons. At ATLAS the hardware
trigger is often called level 0 while at CMS it is often called the level 1 trigger. Higher-level triggers (levels 1 or 2) involve processing with software, such as
jet finding algorithms.  Triggering is extremely important -- if an event does not set off a trigger it is lost forever.

\begin{figure}[t]
\begin{center}
  \includegraphics[width=0.49\textwidth]{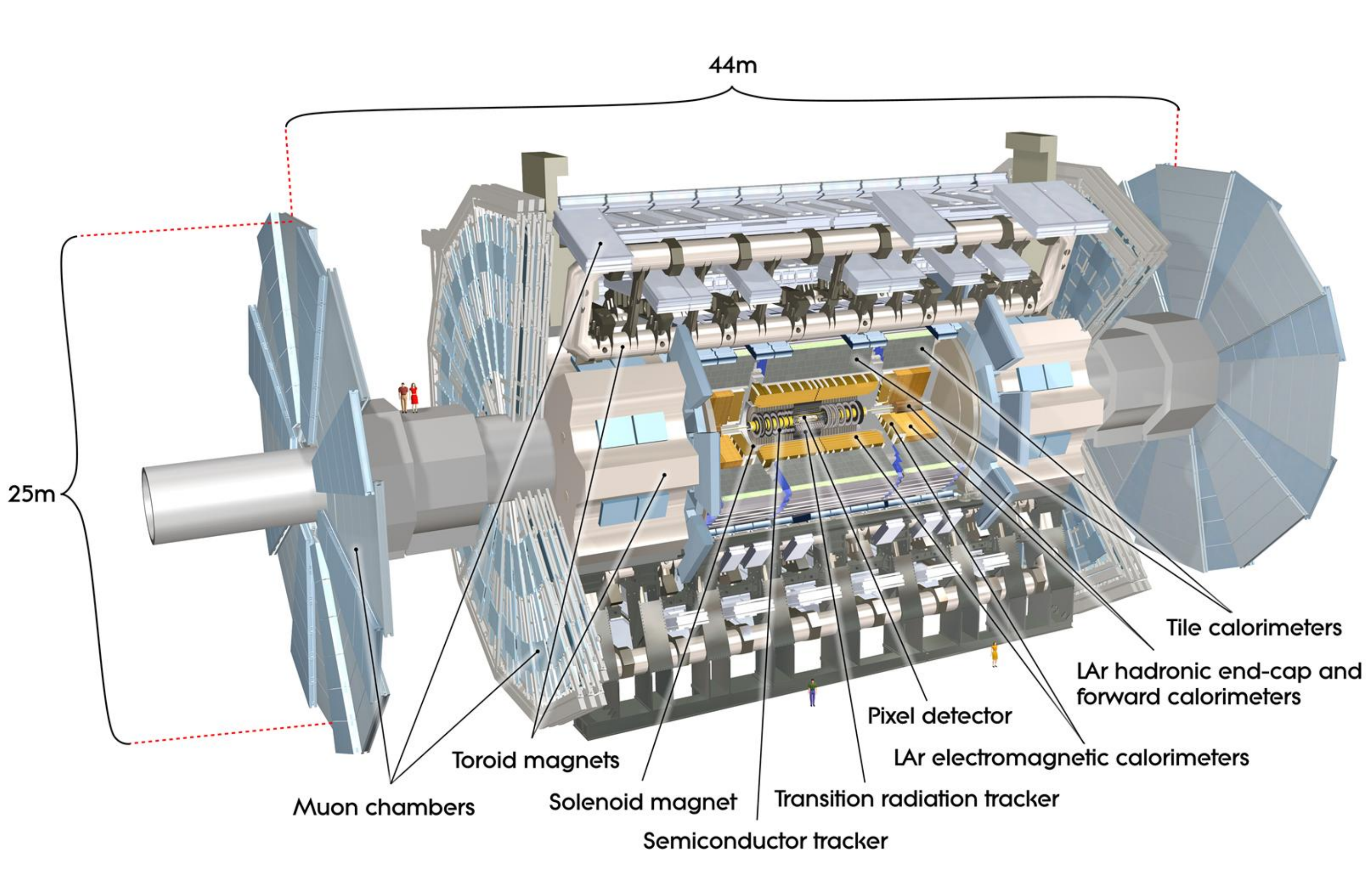}
  \hspace{5mm}
 \includegraphics[width=0.45\textwidth]{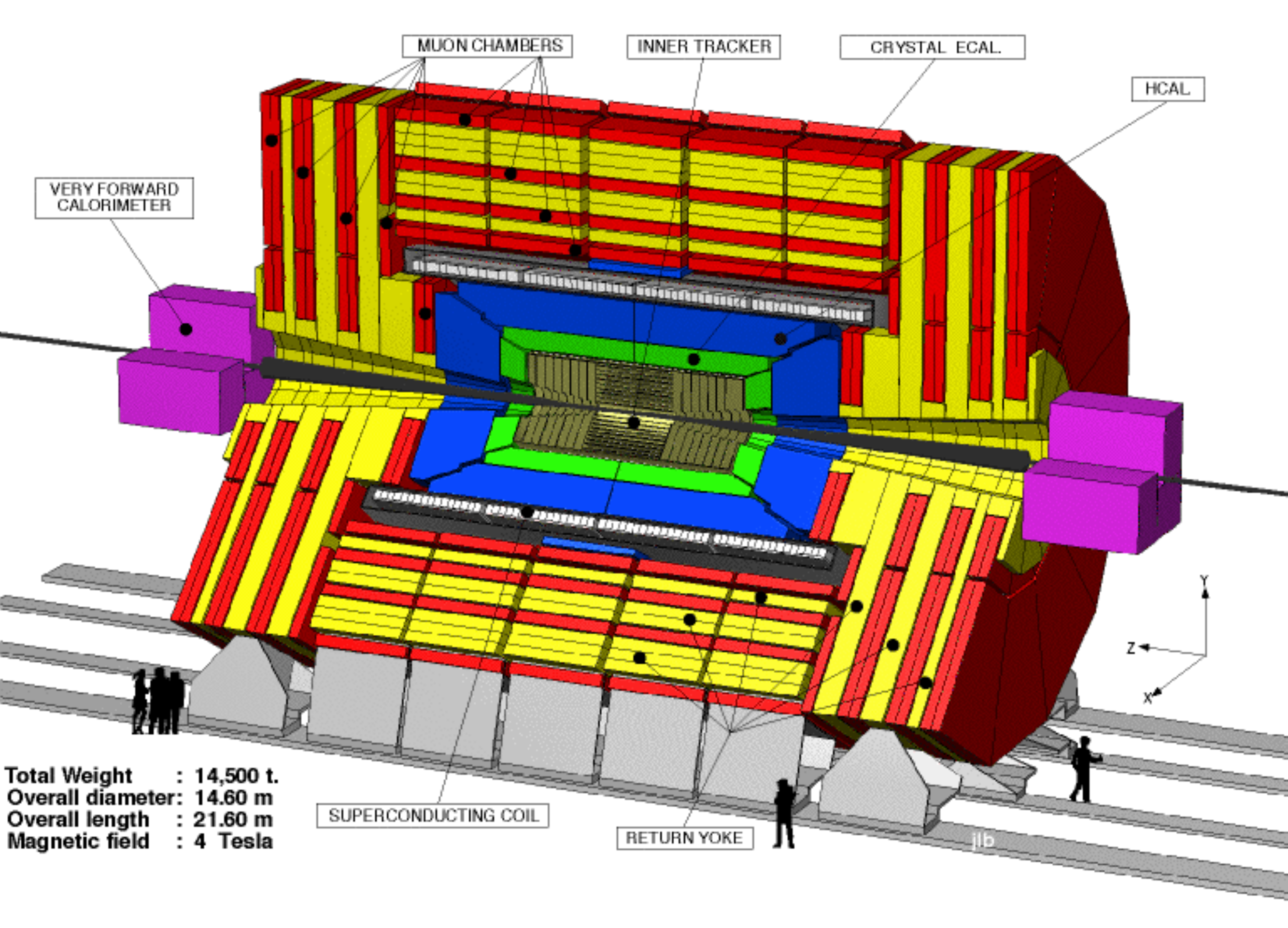}  
  \caption{
The ATLAS and CMS detectors.
  }
  \label{fig:detectors} 
\end{center}
\end{figure}

It's useful to have a rough sense of what the main detectors at the LHC, ATLAS and CMS can measure. Both detectors have the same basic design (see Fig.~\ref{fig:detectors}): an inner detector close to the beam line to measure tracks of charged particles, followed by an electromagnetic  calorimeter (ecal) which measures (mostly) the energy of electrons and photons, then a hadronic calorimeter (hcal) which
measures the energies of neutrons and protons, then at the outside a muon detector. The tracking system in the inner detector
has different parts. For example, on ATLAS, the region closest to the beam line has a set of silicon pixel detectors.
Outside of this is a semiconductor tracker, also made of silicon, followed by a a transition radiation tracker. These systems
have decreasing resolution away from the beam but combine together to give a fantastic picture of what charged particles
were produced by the collision and what direction they went in. At ATLAS, the ecal is made of liquid argon. The hcal is made of
plastic scintillator tiles and iron. On CMS, the ecal is lead tungstate, and the hcal has plastic scintillator and brass. 
CMS is ``compact" with a 4 Tesla magnetic field, while ATLAS has a 2 Tesla field. None of these differences are very important for
phenomenology -- to a good approximation the resolution of the two detectors can be considered the same. 

\section{Kinematics \label{sec:kin}}
The basic picture of a collision at the 13 TeV LHC is as follows. The two protons come in with back-to-back momenta of 6.5 TeV each conventionally taken to be in the $z$ direction. The proton mass is negligible compared to these energies, so we write the proton momenta as
\be
P_1^\mu = (6.5~\TeV, 0, 0, 6.5~\TeV),\qquad
P_2^\mu = (6.5~\TeV, 0, 0, -6.5~\TeV),\qquad
\ee

I like to think of the proton as a big vegetable soup. Most of the soup is broth (the soft gluons), but there are also chunks
of  potatoes or carrots in it (harder gluons, quarks or antiquarks). If you collided two beams of soup against each other,
most of the time, the broth would just scatter, making a big mess (minimum bias events). But sometimes, a potato
will hit against a carrot (hard $qg$ scattering) and you might get a significant amount of splatter going at transverse directions to the beams. It is these collisions we are most interested in. 

When there is a hard scatter, we write the parton momenta as
\be
p_1^\mu = x_1 P_1^\mu, \qquad
p_2^\mu = x_2 P_2^\mu, \qquad
\ee
Here $x_1$ and $x_2$ are the fraction of the proton's momenta in the scattering partons. 
According to the parton model,  $x_1$ and $x_2$ are distributed probabilistically and independently. That is, the probability
of finding momentum fraction $x_1$ in proton one is independent of what happens in proton 2. This independence is an example of factorization,
 key to being able to calculate anything at all at the LHC using perturbation theory. 

\begin{figure}[t]
\begin{center}
\vspace{-20mm}
\begin{tikzpicture}
\node at (0,0) {\includegraphics[width=0.5\columnwidth, trim = {0mm 0mm 30mm 70mm}]{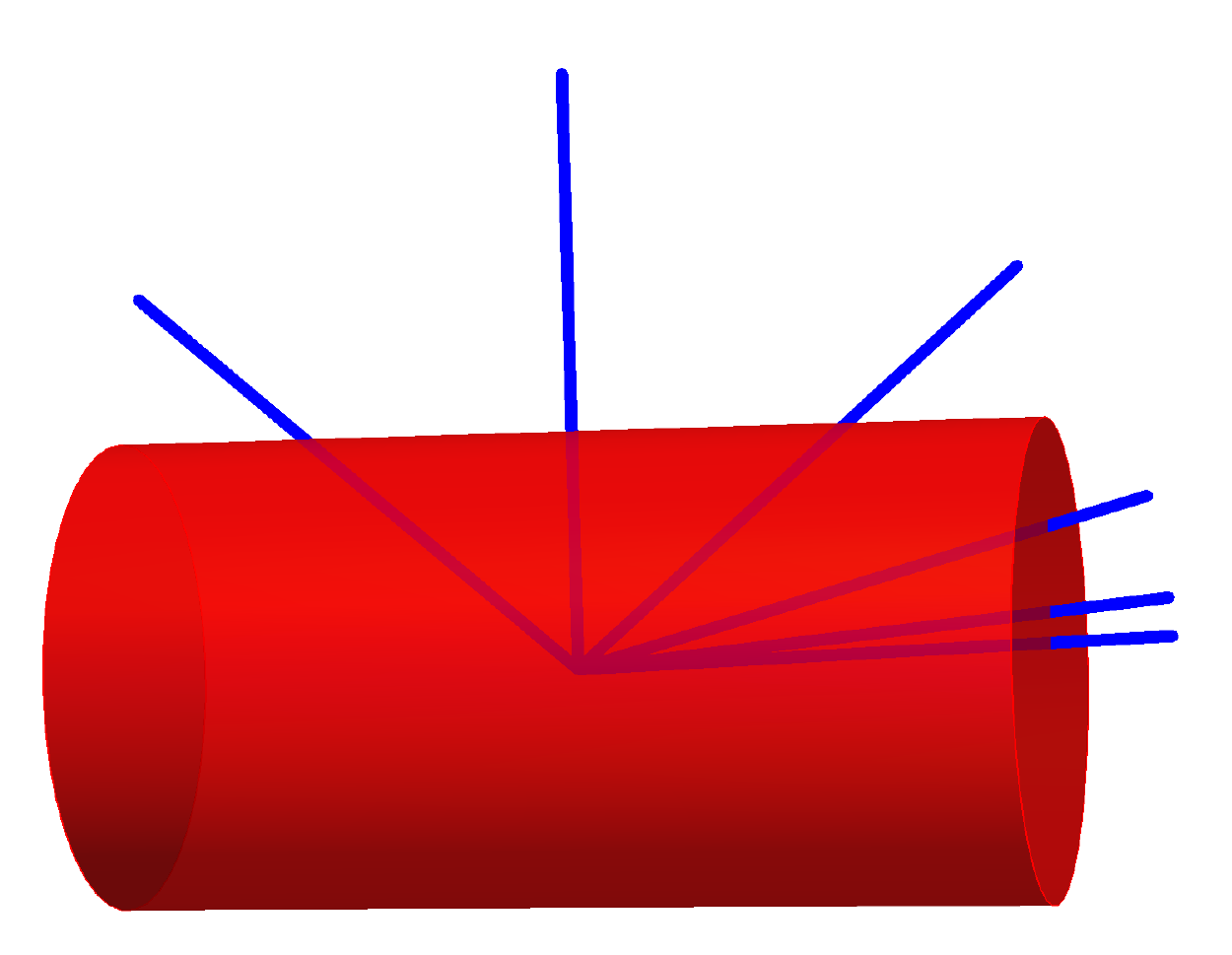}};
\node [above] at (0,6) {$\eta=0$};
\node [above] at (0,5.5) {$\theta=90^{\circ}$};
\node [above] at (3.5,5) {$\eta=1$};
\node [above] at (3.5,4.5) {$\theta=40^{\circ}$};
\node [above] at (-2,5) {$\eta=-1$};
\node [above] at (-2,4.5) {$\theta=130^{\circ}$};
\node [above] at (7.8,2.7) {$\eta=2,~~\theta=15^{\circ}$};
\node [above] at (8,1.9) {$\eta=3,~~\theta=6^{\circ}$};
\node [above] at (8.2,1.3) {$\eta=4,~~\theta=2^{\circ}$};
\node[rotate=0] at (-5,1.5) {$\phi$};
\draw [->,line width=2] (-4.2,3) to [out=-110, in = 110] (-4.2 ,0);
\end{tikzpicture}
  \caption{Pseudorapidity $\eta$ is a geometric quantity. It is a function of polar angle $\theta$ that goes from $\infty$ to $-\infty$ as $\theta$ goes from $0$ to $\pi$.  Azimuthal angle $\phi$ goes around the beam.
  }
  \label{fig:eta} 
\end{center}
\end{figure}
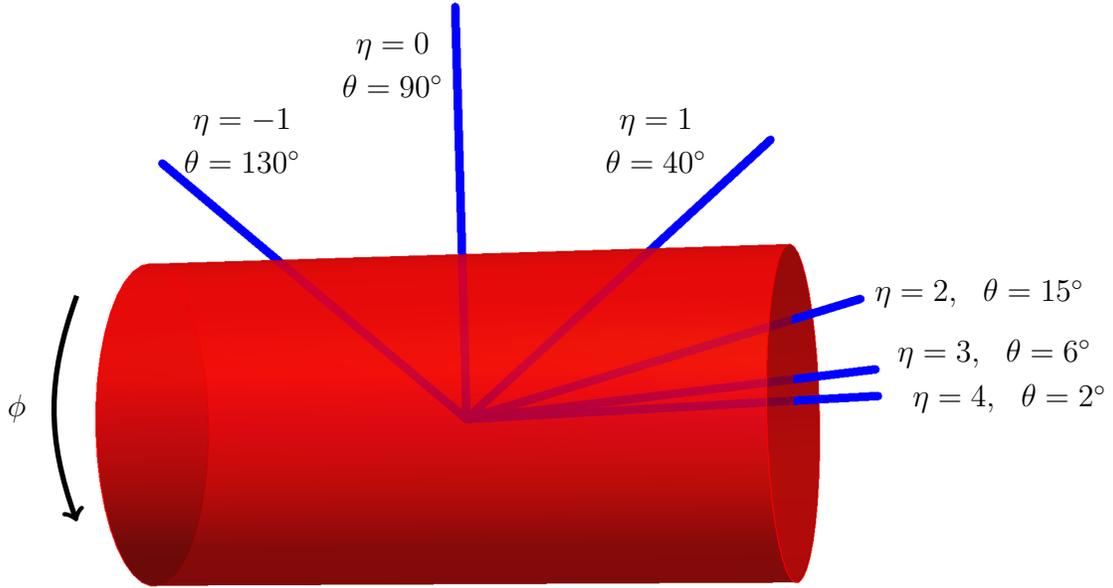

So now we have these partons of momentum $p_1$ and $p_2$ colliding and very high energy. Say they collide to produce a $Z$ boson which then decays into an $e^+/e^-$ pair. We might be interested in the angular separation between these leptons. We can measure the azimuthal angle $\phi$, around the cylinder described by the beam line, and also the polar angle $\theta$
defined with the beams at $\theta =0$ and $\theta = \pi$ and the center of the beams at $\theta = \frac{\pi}{2}$, as in Fig.~\ref{fig:eta}.
 If $(p_1+p_2)^2 = m_Z^2$, the $Z$ is produced at rest in the partonic center of mass frame. However in the lab frame, the $Z$ is not at rest because $p_1+p_2$ can
have some net $z$-momentum, $p_z$. Thus the $e^+$ and $e^-$ pair will be back-to-back in azimuth ($\Delta \phi = \pi$), however
their separation in $\theta$ will depend on this net $p_z$. If $p_z=0$, the leptons will have equal and opposite polar angles. But if  the pair has some net momentum along the $z$ axis they will get closer
together in $\theta$. To see this most clearly, consider very large $p_z$ where  both leptons are close to $\theta =0$. 
In fact, values and differences of  $\theta$ angles tend to tell us more about the parton momenta in the proton than about the angular distribution from $Z$ decays.

Because angles in the lab frame are generally not very interesting from the point of view of the partonic collision,  we like to work with variables which have the same values in the lab and partonic center-of-mass frame. Such variables are longitudinally boost invariant. 
A Lorentz boost along the $z$ direction can be parameterized by a number $\beta$ as
\be
K_z = \begin{pmatrix} 
\cosh \beta & 0 & 0 &\sinh \beta \\
0 & 1 &0 & 0 \\
0 & 0 & 1 & 0 \\
\sinh\beta & 0 & 0 & \cosh \beta
\end{pmatrix}
\ee
A generic momentum $p^\mu = (E, p_x, p_y, p_z)$
transforms under $p \to K_z \cdot p$ as
\begin{align}
E  &\longrightarrow  E \cosh \beta + p_z \sinh \beta \\
p_x &\longrightarrow p_x\\
p_y &\longrightarrow p_y \\ 
p_z &\longrightarrow p_z \cosh \beta + E \sinh \beta
\end{align}
Thus the $x$ and $y$ components of the momentum, $p_x$ and $p_y$, called the {\bf transverse momenta}, are
boost invariant. We often use both the vector and scalar transverse momentum
\be
\vec{p}_T \equiv \left(  p_x, p_y \right), \qquad p_T \equiv |p_T|
\ee
The azimuthal angle
\be
\phi \equiv \tan^{-1} \frac{p_x}{p_y}
\ee
 is also boost invariant.  
 
  To find another boost invariant quantity, let us introduce the shorthand $c \equiv \cosh \beta$ and $s = \sinh  \beta$ so that $c^2 - s^2 = 1$. Then under a boost,
 \be
\frac{E + p_z}{E-p_z} 
\longrightarrow \frac{E(c+s)+p_z(c+s)}{E(c-s) - p_z (c-s)}\times \frac{c+s}{c+s} =\frac{E+p_z}{E-p_z}(c+s)^2
\ee
That is, under a boost this combination rescales by a $\beta_z$-dependent but momentum-independent amount. The log of this combination  therefore shifts under a boost and the difference of logs is independent of $\beta_z$. This motivates defining {\bf rapidity} as
\be
y = \frac{1}{2} \ln \frac{E + p_z}{E-p_z}  
\longrightarrow y + \ln (c+s)
\ee
The {\it difference} of rapidities $y_1$ and $y_2$ for two momenta $q_1$ and $q_2$ is boost invariant. We therefore define
the angular separation as
\be
R = \sqrt{ (\Delta \phi)^2 + (\Delta y)^2}
\ee
This angular separation is boost invariant. Plotting distributions as functions of rapidity rather than polar angle makes it
easier to disentangle the physics of the protons that produced the boost from the physics of the hard collision that
we are studying.

To get intuition for rapidity, consider massless particles. These have $E=| \vec{p} |$.
Then, drawing a little momentum triangle:
\be
\begin{tikzpicture}[scale=1.25]
  \coordinate [label=left:$$] (C) at (-1.5cm,-1.cm);
  \coordinate [label=right:$$] (A) at (1.5cm,-1.0cm);
  \coordinate [label=above:$$] (B) at (1.5cm,1.0cm);
  \draw [->,line width=2] (A) to node[right] {$p_T$} (B) ;
  \draw [->,line width=2] (C) to node[above,rotate=35] {$|\vec{p}|$}  (1.4cm, 0.9cm) ;
  \draw [->,line width=2] (C) to node[below] {$p_{z}$} (A) ;
  \tkzMarkAngle[size=1cm,color=black](A,C,B)
\node [above] at (-0.2cm,-0.8cm) {$\theta$};
\end{tikzpicture}
\ee
 we see that $\cos\theta = \frac{p_z}{|\vec{p}|} = \frac{p_z}{E}$. So,
\be
y = \frac{1}{2} \ln \frac{E + p_z}{E-p_z}   =  \frac{1}{2} \ln \frac{1 +  \cos \theta}{1 - \cos \theta} 
= \frac{1}{2} \ln \frac{2\cos^2\frac{\theta}{2}}{2 \sin^2\frac{\theta}{2}} = \ln \cot \frac{\theta}{2}, \qquad m=0~\text{only}
\ee
Thus there is a simple mapping between rapidity and angle for massless particles.
 This motivates defining
{\bf pseudorapidity} $\eta$ as
\be
\eta \equiv \ln \cot \frac{\theta}{2}
\ee
Taylor expanding this relation around $\theta \approx \frac{\pi}{2}$, we see
\be
\eta \approx \frac{\pi}{2} - \theta 
\ee
 For example
\be
\begin{array}{|c||c|c|c|c|}
\hline
\theta-\frac{\pi}{2} & 0 &  0.1 & - 1 & 10^{\circ}~ \text{from beam}\\
\hline
\eta &  0  & - 0.1001 & 1.22 & \pm2.5\\
\hline
\end{array}
\ee
Some values for $\theta$ and $\eta$ are shown in Fig.~\ref{fig:eta}. Note that particles at $\eta \sim \pm 5$ are practically down the beam line. The ATLAS and CMS detectors measure particles up to pseudorapidities of around $\pm 5$. 

In summary, rapidity is a {\it kinematic quantity} defined as $y \equiv \frac{1}{2} \ln \frac{E + p_z}{E-p_z}$. Rapidity itself is
not boost invariant, but differences in rapidity are boost invariant. 
Other boost invariant quantities are
$\vec{p}_T \equiv \left(  p_x, p_y \right)$ and $\phi = \tan^{-1} \frac{p_x}{p_y}$. 
Pseudorapidity $\eta \equiv  \ln \cot \frac{\theta}{2}$
is a {\it geometric quantity}. It is equal to rapidity only for massless particles. For massive particles, differences in pseudorapidities are not boost invariant. 

\section{Observables \label{sec:obs}}
All collider observables are functions of the momentum and energy of the particles produced. 
Ideally, we would like to measure the 4-momentum  of every particle in the event, but this is not quite possible in practice. To a first approximation, what can be measured is the energy of all the particles stable on detector timescales, through
deposits to the various calorimeters, and their directions ($\eta, \phi)$. Since most particles of interest are essentially massless, energy and angle are enough to reconstruct the momentum. 
Most particles deposit all their energy within the central calorimetry system. The two exceptions are neutrinos, which leave the detector rarely having interacted at all, and muons.
Muons are like little cannonballs that get all the way through the detector before depositing all their
energy. Nevertheless, the momentum of muons can be measured by using the curvature of their trajectories through the muon calorimeter. To do so requires strong magnetic fields to shift the muons' trajectories, and lots of space to see the small curvature of the energetic tracks. The
muon system is much of the reason why ATLAS and CMS are so big. Curvature of tracks is also used in the inner detector, to distinguish charged particles like electrons, positrons and pions, and to help measure their 3-momentum. 

A standard observable constructed from the particles' momenta is {\bf missing transverse momentum}:\be
\ptmiss \equiv - \sum_j \vec{p}_T
\ee
Missing transverse momentum is a 2-vector. A related quantity is missing transverse energy
\be
\met \equiv | \ptmiss |
\ee
Missing transverse energy (MET) is a scalar. 

For example, if an event has a $W^-$ boson in it which decays to  $e^- \nu$, then we will only see the electron,
not the neutrino. The electron's 4-momentum can essentially be measured. The momentum of the neutrino
should have opposite $p_x$ and $p_y$ components to the electron, but its $p_z$ component does not have
to be opposite to that of the electron, due to the longitudinal boost of the partonic system. Thus $\ptmiss$ gives the transverse components
of the neutrino. Knowing that the $W$ boson was on-shell gives an additional constraint that allows the neutrino momentum
to be fully reconstructed (up to a 2-fold ambiguity from the 2 roots of the quadratic equation $m_W^2 = p_e^2 + p_\nu^2$). 
If there is more than one neutrino in the event, we cannot reconstruct all the neutrinos' momenta. 
For example, in $p \to Z\to \nu \nu$ events, there may be no transverse momentum at all, so the neutrinos could have gone anywhere (of
course, there is nothing at all to see here so this event would not even trigger). 

Another quantity commonly discussed is $H_T$. There is not a unique definition of $H_T$, but it usually refers to the scalar
sum of the missing transverse momentum in certain objects.  For example, we might see
\be
H_T = \left|\sum_{\text{jets}~j} \vec{p}_T^{~j} \right|,\qquad \text{or} \qquad 
H_T = \sum_{\text{jets}~j}\left| \vec{p}_T^{~j}  \right|,\qquad \text{or} 
\qquad 
H_T = \sum_{\text{leptons}~j}\left| \vec{p}_T^{~j}  \right|,\qquad 
\ee
Whenever you see $H_T$ used, makes sure you know what quantity it refers to.

We also are often interested in the invariant mass of some objects
\be
m_{\text{objects}} = \left| \sum_{\text{objects}~j} p^\mu_j \right|^2
\ee
For example, in $ pp \to \gamma \gamma$ events, we might look at the invariant mass of the two photons. Plotting 
the number of events observed as a function of invariant mass should show a resonant peak at 
the mass of the Higgs boson,  $m_{\gamma\gamma} \sim m_h$. 

The invariant mass of two particles is $m^2 = \sqrt{ (E_1 + E_2)^2 - (\vec{p}_1 + \vec{p}_2)^2}$. Sometimes we don't
know all three components of the momentum, so the best we can do is consider the {\bf transverse mass}
\be
m_T \equiv \sqrt{(E_T^1 + E_T^2)^2 - (p_T^1 + p_T^2)^2}
\label{mTdef}
\ee
where $E_T = \sqrt{m^2 + p_T^2}$ is the {\bf transverse energy}. 
If both $p_1$ and $p_2$ are purely transverse ($\eta=0$), then $m_T = m$. If $p_1$ and $p_2$ are purely longitudinal then $m_T=0$. For other situations, $0< m_T < m$. It is helpful sometimes to expand
\be
m_T = \sqrt{ (E_T^1)^2+(E_T^1)^2+2(E_T^1)(E_T^2)}
\ee
The usefulness of transverse mass is discussed more in the next section. 

\section{Distributions \label{sec:dist}}
Consider the production of $e^+ e^-$ pairs through an intermediate $Z$ boson. This process is called the {\bf Drell-Yan Process} and is one of the simplest and most important processes that occur at hadron colliders. One thing we might look
at in Drell-Yan is the invariant mass of the lepton pair: $m^2 = (q_1 + q_2)^2$ where $q_1$ and $q_2$ are the lepton momenta.
If we measure this observable, it should have a peak at the $Z$-boson mass. One nice feature of Drell-Yan is that if we
only measure the lepton momenta, the cross section has been proven to factorize. That is, we can write it as
\be
\frac{d\sigma}{dm^2} = \int d x_1 d x_2 f_{q}(x_1,\mu) f_{\bar{q}}(x_2,\mu) \frac{d \hat{\sigma}(q\bar{q} \to e^+ e^-,\mu)}{d m^2}
\label{dyfact}
\ee
Here $x_1$ and $x_2$ are the momentum fractions of the quark and anti-quark partons in the protons and $f_q(x_i,\mu)$ are the parton distribution functions which depend on a factorization scale $\mu$. This factorization formula as written is really
only valid at leading order. At next-to-leading order, one must sum over other production channels and the final state can
have more particles in it, for example, a gluon radiating off of the initial state quarks. 
%

We use $\hat{s}$ to denote the partonic center-of-mass energy. In general, we put hats on quantities when they are partonic.
This $\hat{s}$ can be easily related to the machine center-of-mass energy $S= (13\, \text{TeV})^2$ by using
\be
\hat s = (p_1 +p_2)^2  = p_1^2 + p_2^2 + 2 p_1\cdot p_2 = 2 x_1 x_2 P_1 \cdot P_2 = x_1 x_2 S
\ee
where $p_1$ and $p_2$ are the (massless) parton momenta. Because of this relation, the cross section
depends only on the combination $x_1 x_2$ and not on $x_1$ and $x_2$ separately. Thus
it is helpful to write  Eq.~\eqref{dyfact} as
\be
\frac{d \sigma}{d m^2} = \cL_{q\bar{q}} (m^2) \frac{d \hat{\sigma}}{d m^2}
\ee
where the quark-antiquark {\bf luminosity function} is defined by
\be
\cL_{q\bar{q}}(\hat{s}) \equiv \int d x_1 d x_2 f(x_1,\mu) f(x_2,\mu) \delta( x_1 x_2 S - \hat{s})
\ee
The nice thing about Drell-Yan is that by measuring the lepton invariant mass one is directly measuring the partonic $\hat{s}$ (at leading order). 

This factorization of the cross section into a luminosity function times a partonic cross section is very useful.  By looking at the values of $ \cL_{q\bar{q}} (\hat s) $ at
various energies, we can get a sense of the relative importance of different partonic channels. Note that the cross
section is  the product of $\cL_{q\bar{q}} (\hat{s})$ and a partonic cross section only if there is only $s$-channel production, as for Drell-Yan at leading order. If there is also a $t$-channel contribution, the cross section will depend on $x_1$ and $x_2$ in some
combination other than $x_1 x_2$ and luminosity functions are not so useful.

So what about $\frac{d \sigma}{d m^2}$? At leading order, it is given by one Feynman diagram
\be
\frac{d \hat \sigma}{d \hat{s}} \sim 
\left|
\begin{gathered}
\resizebox{20mm}{!}{
     \fmfframe(0,0)(0,0){
\begin{fmfgraph*}(30,20)
\fmfstraight
	\fmfleft{L1,L2}
	\fmfright{R1,R2}
	\fmf{fermion}{L1,v1,L2}
	\fmf{fermion}{R1,v2,R2}
	\fmf{photon,label=$Z$,l.s=right}{v1,v2}
	\fmfv{label=$g$,l.a=60}{v1}
	\fmfv{label=$g$,l.a=120}{v2}
\end{fmfgraph*}
}}
\end{gathered}
\right|^2 \sim \left| \frac{g^2}{\hat{s} - m_Z^2 + i \Gamma m_Z}\right|^2
=\frac{g^4}{(\hat{s} - m_Z^2)^2  + \Gamma^2 m_Z^2}
\ee
This determines the shape of the distribution: it has the Breit-Wigner form, with a peak at $\hat{s} = m_Z$ and width $\Gamma_Z$. The luminosity functions are smooth over the support of the partonic cross section, thus the full measured distribution
will also have the Breit-Wigner shape. Measuring $\frac{d \sigma}{d m^2}$ can therefore be used to measure both the $Z$ boson mass
and its width.

The width $\Gamma_Z$ can also be computed by Feynman diagrams
\be
\Gamma_Z \sim  \sum_{\text{decay modes}}
\left|
 \begin{gathered}
\resizebox{20mm}{!}{
     \fmfframe(0,0)(0,0){
\begin{fmfgraph*}(30,20)
\fmfstraight
	\fmfleft{v1}
	\fmfright{R1,R2}
	\fmf{fermion}{R1,v2,R2}
	\fmf{photon,label=$Z$,l.s=right}{v1,v2}
	\fmfv{label=$g$,l.a=120}{v2}
\end{fmfgraph*}
}}
\end{gathered}
\right|^2 \sim g^2
\ee
As the coupling goes to zero, the width also goes to zero. In general, for a perturbative theory (small couplings), widths are small compared to the mass.\footnote{
For strongly coupled theories like QCD, widths may not be small. For example, bound states of gluons called glueballs generally have widths of order their mass. These widths are so large that
it is not even clear if glueballs should be considered particles.}
 For example, the
width of the $Z$ boson is  $\Gamma_Z =2.5$ GeV is much less than $m_Z =91$ GeV.
In the limit that $\Gamma_Z/m_Z \to 0$, the Breit-Wigner distribution reduces to a $\delta$-function distribution:
\be
\frac{d \hat\sigma}{d \hat{s}} 
\sim \frac{g^4}{(\hat{s} - m_Z^2)^2  + \Gamma^2_Z m_Z^2} \to g^4\frac{\pi}{\Gamma_Z m_Z} \delta(\hat{s} - m_Z^2)
\label{narrow}
\ee
Note that although the cross section scales like $g^4$, since $\Gamma_Z \sim g^2$, on resonance the cross section only scales like $g^2$. This is a resonant enhancement: the cross section is {\it much larger} on resonance than off resonance, by a factor of $\frac{1}{g^2}$. 

Because of the $\delta$-function in Eq.~\eqref{narrow}, the cross section nicely factorizes into production and decay:
\be
 \hat \sigma(q \bar{q}  \to Z \to e^+ e^-)   = \hat \sigma( q\bar{q} \to Z) \times \text{BR} (Z \to e^+e^-)
 \ee
 where the {\bf branching ratio} is defined as
 \be
 \text{BR} (Z \to e^+e^-)=   \frac{\Gamma(Z \to e^+e^-)}{\Gamma_Z} 
\ee
Here $\Gamma_Z = 2.5$ GeV is the total decay width of the $Z$ boson while
 $\Gamma(Z \to e^+e^-) \approx 0.03~ \Gamma_Z$ is the partial width for the $Z$ to decay to $e^+ e^-$ pairs.
 The factorization into production and decay follows from this {\bf narrow-width approximation}. The narrow width
 approximation says that quantum-mechanical interference between corrections from production and decay can
 be neglected so that production and decay are separate well-defined probabilities. 
 
At least one of the branching ratios is usually of order $1$. So while the original process scaled generically like $g^4$, 
the actual rate is determined by  $\sigma(q \bar{q}  \to Z \to e^+ e^-)$ which scales only like $g^2$. As noted
above, the missing factors of $g^2$ are canceled by the resonance enhancement.

Next, consider the hadroproduction of a $W$ boson which decays to $e^- \nu$. We would love to be able to measure the $W$ mass from looking at the $e^- \nu$ invariant mass, like we do for the $Z$ boson. However, this is impossible as the neutrino is unobservable. Instead, we can look at the electron alone. It is helpful to define $\theta^\star$ as the polar angle of the electron in
the $W$ center-of-mass frame. Then a basic QFT calculation gives
\be
\frac{1}{\sigma_0} \frac{d \hat \sigma(q \bar{q} \to W \to e^- \nu)}{d \cos \theta^\star} = \frac{3}{8} (1+ \cos^2 \theta^\star)
\label{csd}
\ee
where $\sigma_0$ is the total leading-order cross section. 
Of course, one cannot directly measure $\theta^\star$, since one does not know the $W$ boson rest frame, as 
the longitudinal momentum $q_z$ of the neutrino is unknown. However, for a given $W$ mass, we get an extra constraint
which can be used to determine $q_z$ and hence $\theta^\star$. Indeed, in the $W$ rest frame, 
$p_W^\mu = (m_W, \vec{0} )$. Then, treating the electron and neutrino momenta as massless,
 the electron momentum is $p_e^\mu  = (\frac{m_W}{2}, \vec{p})$ and
the neutrino momentum is $p_\nu^\mu = (\frac{m_W}{2}, - \vec{p})$ for some 3-vector $\vec{p}$.
Drawing a little triangle
\be
\begin{tikzpicture}[scale=1.25]
  \coordinate [label=left:$$] (C) at (-1.5cm,-1.cm);
  \coordinate [label=right:$$] (A) at (1.5cm,-1.0cm);
  \coordinate [label=above:$$] (B) at (1.5cm,1.0cm);
  \draw [->,line width=2] (A) to node[right] {$p_T$} (B) ;
  \draw [->,line width=2] (C) to node[above,rotate=35] {$|\vec{p}\,|=\frac{m_W}{2}$}  (1.4cm, 0.9cm) ;
  \draw [->,line width=2] (C) to node[below] {$p_{z} = \sqrt{(\frac{m_W}{2})^2 - p_{T}^2}$} (A) ;
  \tkzMarkAngle[size=1cm,color=black](A,C,B)
\node [above] at (-0.2cm,-0.8cm) {$\theta^\star$};
\end{tikzpicture}
\ee
we see that $p_{z} = \sqrt{(\frac{m_W}{2})^2 - p_{T}^2}$ and 
\be
\cos \theta^\star = \frac{p_z}{|\vec{p}\,|} = \sqrt{1-\frac{4 p_{T}^2}{m_W^2}}
\ee
In other words, knowing $m_W$ lets us trade $\theta^\star$ for $p_{T}$. 

\begin{figure}[t]
\begin{center}
\begin{tikzpicture}
\node at (-2,0) {\includegraphics[width=0.5\columnwidth]{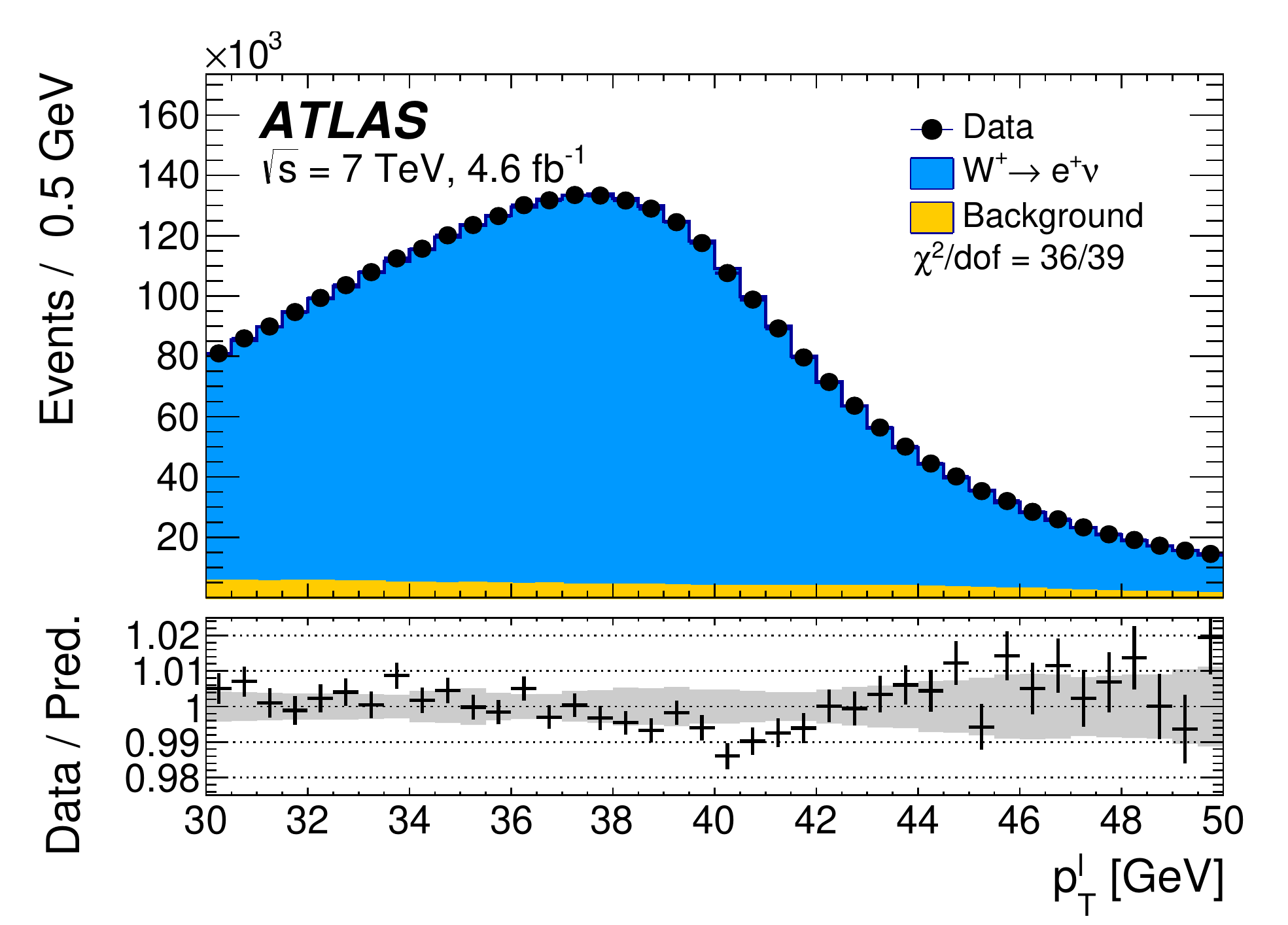}};
\node at (7,0) {\includegraphics[width=0.5\columnwidth]{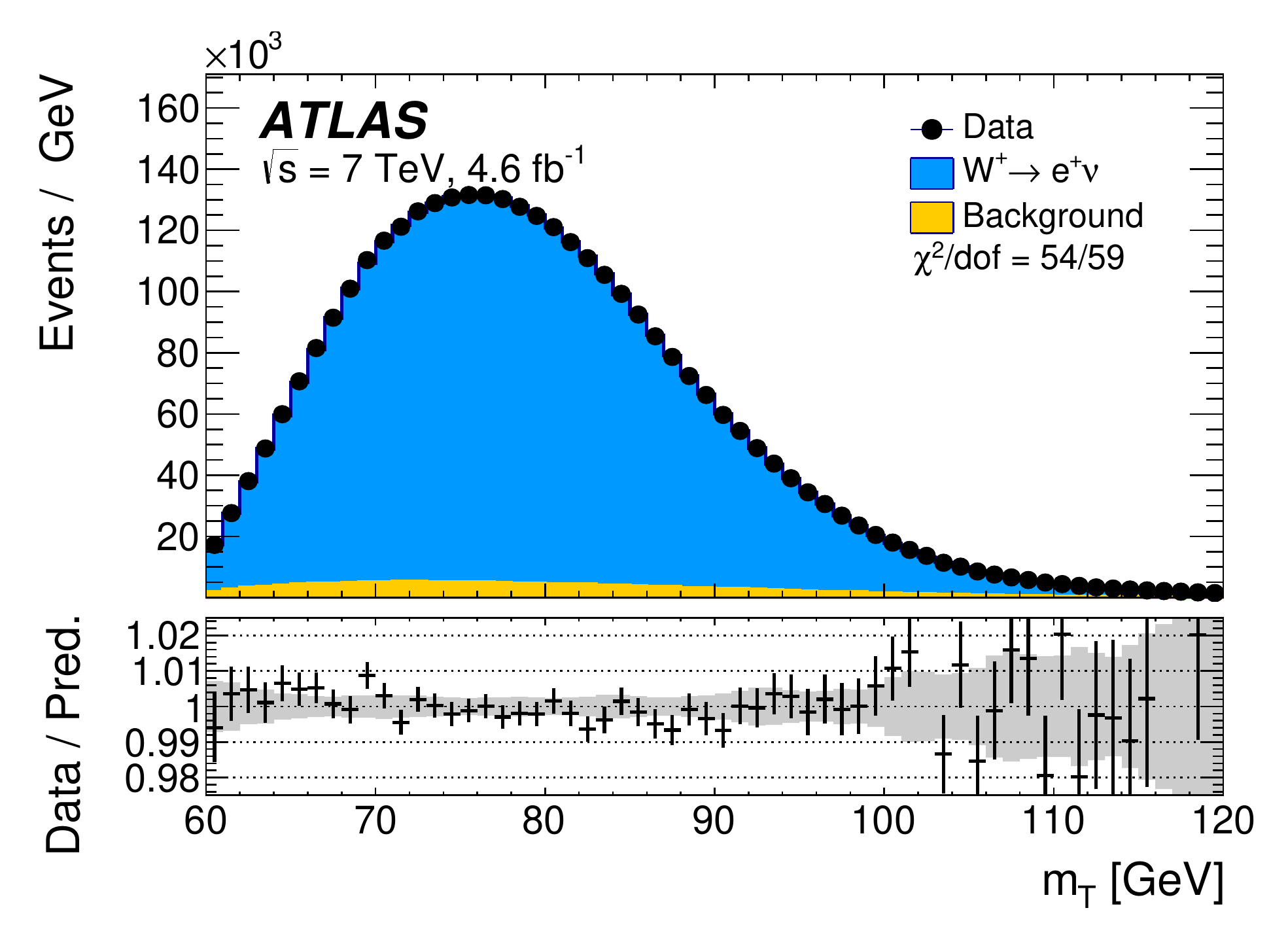}};
\draw [->,black!30!red, line width = 2] (-1,3) node[above, black!30!red] {Jacobian peak at $\sim \frac{m_W}{2}$} -- (-2,2) ;
\draw [->,black!30!red, line width = 2] (7,3) node[above, black!30!red] {Jacobian peak at $\sim m_W$} -- (6.3,2.3) ;
\end{tikzpicture}
\end{center}
\caption{Distribution of the lepton $p_T$ and transverse mass $m_T$ for  $W^+ \to e^+ \nu$ events and their background at the LHC using 7 TeV data. Figure from ATLAS~\cite{Aaboud:2017svj}. 
\label{fig:mt}}
\end{figure}

Changing variables in Eq.~\eqref{csd} from $\theta^\star$ to $p_T$ gives
\be
\frac{1}{\sigma_0} \frac{d \hat \sigma(q \bar{q} \to W \to e^- \nu)}{d p_T} =
\underbrace{\frac{3}{m_W^2} \frac{p_T}{\sqrt{1-\frac{4 p_T^2}{m_W^2}} }}_{\text{Jacobian factor}}\underbrace{\left(1-\frac{2 p_T^2}{m_W^2}\right)}_{1+\cos^2\theta^\star}
\ee
This change of variables has an interesting effect. While the $\cos\theta^\star$ distribution was
a simple polynomial in $\cos \theta^\star$, the $p_T$ distribution seems to blow up as $p_T \to \frac{m_W}{2}$.
In reality, it does not blow up since higher order effects give the $W$ boson some $p_T$ and resolve the singularity.
Instead, the leading-order singularity shows up as a peak in the distribution at $p_T=\frac{m_W}{2}$. This is known as a  {\bf Jacobian peak}. Measuring the location of this peak is one way the $W$ boson mass is measured. You can see the Jacobian peak in the data in Fig.~\ref{fig:mt}

It turns out the lepton $p_T$ is not the best observable to use to find the $W$ mass. One reason is that the $p_T$ distribution is very sensitive to higher order effects. For example, the NLO effects are not small on the height of the peak -- they take it from infinity to something finite. Moreover, no information about the neutrino is being used. One naturally would expect that a better $W$ mass measurement should result from incorporating the $p_T$ of the neutrino, which is known from the missing $p_T$ in the event, than by throwing it out. 
In $W \to e\nu$ decays, the $W$ boson mass is
\be
m_W^2 = (\underbrace{E_e}_{\text{known}} +\underbrace{ E_\nu}_{\text{unknown}})^2
 - (\underbrace{\vec{p}_{Te}}_{\text{known}} + \underbrace{\vec{p}_{T \nu}}_{\text{known}})^2 
 - (\underbrace{p_{ze}}_{\text{known}}  + \underbrace{p_{z\nu}}_{\text{unknown}})^2
\ee
There are two unknowns here.  We can use the neutrino's mass-shell condition to remove one unknown, but there is still one unknown quantity. That is to say, we cannot reconstruct the $W$ mass exactly. However, note that if we look at the transverse mass from Eq.~\eqref{mTdef}:
\be
m_T^2 = (E_{eT} + E_{\nu T})^2 - (\vec{p}_{Te} + \vec{p}_{T\nu})^2
\ee
where the {\bf transverse energy} for a particle is defined as
\be
E_T^2 = m^2 + p_T^2
\ee
we see that everything is known. Moreover,  the transverse mass, in contrast to the lepton $p_T$, depends on neutrino transverse momentum so it incorporates more information.

To get a sense of what the transverse mass is, consider the situation when the electron and neutrino are completely transverse,
so $p_{z e } = p_{z \nu} = 0$. Then $E_{eT} = E_e$ and $E_{\nu T} = E_\nu$ and $m_T= m$. Thus for transverse production,
the transverse mass reduces to the invariant mass. 

In general, the $W$ boson will be produced with some longitudinal momentum. The lab frame is related to the $W$ rest frame by a longitudinal boost. This boost doesn't change $m_T$ or $m$, as both are boost invariant. In the $W$ rest frame, $p_e^\mu = (E,\vec{p}_T, p_z)$ and 
$p_\nu^\mu = (E , -\vec{p}_T , -p_z)$. So $m^2 = (2E)^2 = 4 p_T^2 +4 p_z^2$ and 
\be
m_T^2 = (E_{eT} + E_{\nu T})^2-(\vec{p}_{Te} + \vec{p}_{T\nu})^2
=2 E_{eT} E_{\nu T} - 2\vec{p}_{Te}\cdot \vec{p}_{T\nu}
 = 4 p_T^2
\ee
hence $m^2 = m_T^2 + 4 p_z^2$. Thus $m_T \le m$. 
We conclude that (at leading order) the transverse mass can never be larger than the $W$ boson mass. 

Including the Jacobian factor, the distribution in $m=(q_e + q_\nu)^2$ and $m_T$ is
\be
\frac{1}{\sigma_0} \frac{d \hat \sigma(q \bar{q} \to W \to e^- \nu)}{ d m^2 d m_T^2} \propto
 \underbrace{ \frac{\Gamma_W m_W}{ (m^2 - m_W^2)^2 - \Gamma_W^2 m_W^2}}_{\text{Breit-Wigner}}
 \underbrace{ \frac{1}{m\sqrt{m^2-m_T^2}}}_{\text{Jacobian}}
 \ee
 From this formula we see that the kinematic endpoint at $m_T = m_W$ has turned
 into a Jacobian peak. As with the lepton $p_T$ due to higher order effects (additional radiation), 
 it is possible to have $m_T > m_W$. The $m_T$ distribution from ATLAS data is shown in Fig.~\ref{fig:mt}.  Both $p_T$ and $m_T$ are useful for measuring the $W$ boson mass. 
  
The power of transverse mass comes partly from its natural leading-order kinematic endpoint at $m_T = m_W$. This endpoint is a leading-order kinematic feature and does not depend on the 
neutrino being massless. If the neutrino were massive (think neutralino in supersymmetry) then the transverse mass would be
\be
m_T(\vec{p}_e, \vec{p}_\nu,m_e, m_\nu) = m_e^2 + m_\nu^2+2 (E_{Te} E_{T\nu}-\vec{p}_{Te} \cdot \vec{p}_{T\nu})
\ee
This is a useful quantity as long as there is only one unmeasured particle, since then $\vec{p}_{T\nu}$ can be reconstructed. But consider
the case when there are two neutrinos, like in $W^+ W^-$ production. Or consider the historically important supersymmetric scenario in which two sleptons $\tilde{l}$ are produced, which subsequently
 decay to electrons and a neutral fermion $\chi$:
$pp \to \tilde l  \tilde l \to e^- \chi e^+ \chi$. 
Think of this like $W$ pair production with the slepton representing the $W$ and the neutralino representing a massive neutrino. 
In this case, not only do we not know the neutralino mass, but because of the two neutralinos, we don't know the transverse momentum of each, only their sum. In such a situation, a useful quantity is
{\bf MT2}:
\be
m_{T2}(m_\chi) \equiv \min_{  \vec{p}_{T\chi1} + \vec{p}_{T\chi2} = \slashed{\vec{p}_T}
}
\Big[
m_T(\vec{p}_{e1},\vec{p}_{\chi 1},m_e,m_\chi),
m_T(\vec{p}_{e2},\vec{p}_{\chi 2},m_e,m_\chi)
 \Big]
\ee
For a putative value of $m_\chi$ one can measure the distribution
of $m_{T2}$. If $m_\chi$ is chosen correctly, there will be an endpoint of the distribution at
$m_{T2} = m_{\tilde{l}}$ (like how $m_T$ has an endpoint at $m_W$ for the correct neutrino mass).
If the neutralino mass is not correct, then there will generically still be an endpoint of the distribution, but it may not
be at the correct slepton mass. However, we can compute
\be
m_{T2}^{\text{max}}(m_\chi)
=\max_{\text{events}} m_{T2}
\ee
This $m_{T2}^{\text{max}}$
has a remarkable feature that it has kink when $m_\chi$ is chosen correctly. Thus
one can in principle use $m_{T2}$ to determine both the neutralino mass, through the location of the kink
in $m_{T2}^{\text{max}}$, and the slepton mass,
through the endpoint.  See~\cite{Lester:1999tx} for more information.

\section{Jets \label{sec:jets}}
The physics of jets is a huge subject and there are many great reviews (e.g. \cite{Salam:2009jx,Altheimer:2012mn}). I will just give here the briefest of overviews. 

A jet is a collection of particles that go towards the same direction in the detector. Intuitively, the come from bremsstrahlung (showering) and
fragmentation of a primordial hard quark or gluon. In practice, jets must be defined from experimental observables, namely
the 4-momenta of the observed particles in the event. There is no unique jet definition. One desirable 
property of a jet definition is that it allows for 
a calculation done at the parton level (quarks and gluons) to agree with a measurement done at the particle level (pions and protons). Another
desirable feature is that a jet algorithm should be local in the detector -- not pulling in particles from far away. This property
is important from an experimental point of view to reduce systematic uncertainty of jet measurements.

The simplest way to define a jet is to draw a cone of size $R=\sqrt{(\Delta \eta)^2 + (\Delta \phi)^2}$ around some 
particle and include everything in that cone as the jet. Such a procedure depends strongly on which particle
is chosen as the center of the cone. For example, one might take the hardest particle in the event as the center. However,
such a choice is not infrared safe: if a particle happened to split into two collinear particles with half its energy in each, then
the hardest particle quickly becomes not the hardest. There are infrared safe cone algorithms, the most notable is
perhaps SIScone, but they are not commonly used.

Most popular jet algorithms are based on the idea of iterative clustering:
\begin{enumerate}
\item Calculate the pairwise distance $d_{ij}$ between every pair of objects.
\item Merge the two closest particles.
\item Repeat until no two particles are closer than some given $R$.
\end{enumerate}
Such an algorithm will result in $n$ jets of size $R$. 

The ``objects" on which the jet algorithm operates can be quarks and gluons (for theory calculations), they can be all measured particles (pions, protons, electrons, etc.). Or they can be energy deposits in the calorimeters or topoclusters (aggregate energy deposits in the ATLAS detector) or particle-flow candidates, as reconstructed by CMS. 

The distance measure can be chosen in different ways. Some options are
\begin{itemize}
\item {\bf JADE} algorithm: $d_{ij} = \frac{m_{ij}}{Q^2}$, where $m_{ij}$ is the invariant mass of the two objects. 
\item {\bf Cambridge/Aachen} algorithm $d_{ij} = R_{ij} = \sqrt{(\eta_i -\eta_j)^2 + (\phi_i - \phi_j)^2 }$. 
\item {\bf $k_T$}: $d_{ij} = \min ( p_{Ti}^2, p_{Tj}^2)\frac{R_{ij}^2}{R^2}$
\item {\bf anti-$k_T$}: $d_{ij} = \min (\frac{1}{p_{Ti}^2},\frac{1}{p_{Tj}^2})\frac{R_{ij}^2}{R^2}$
\end{itemize}
These last 3 measures are all of the form
\be
d_{ij} = \min(p_{Ti}^\alpha, p_{Tj}^\alpha)\frac{(R_{ij})^2}{R^2}
\ee
with $\alpha = 2$ for $k_T$, $\alpha=0$ for Cambridge/Aachen,  and $\alpha=-2$ for anti-$k_T$. 
For hadron collisions, one must also define a distance measure to the beam. For anti-$k_T$, the 
beam distance is
\be
d_{iB} = \frac{1}{p_T^2}
\ee

Once jets have been found, there are different things we can do with them. The primary use of jets relies only on the jet's 4-momentum, defined as the sum of the 4-momenta of all the particles in the jet. This jet 4-momentum should be similar to the 4-momentum of the parton which initiated the jet. Of course, this cannot be exactly true, since the parton momentum is a theoretical construction -- quantum effects necessarily make the quark-radiating-gluons semi-classical picture imprecise. The jet 4-momentum is nevertheless very useful: angular and $p_T$ distributions of jets provide signatures of SM and BSM physics. 

Before the LHC, essentially all that jets were used for was approximating the momentum of a primordial parton. However, in recent years there has been an explosion of interest in {\bf jet substructure}. This revolution in jet physics is due partly to the improved resolution of the detectors
in ATLAS and CMS compared to those of previous machines. With increased angular resolution one can approach a complete classification of all the particles in the jet, rather than just the aggregate energy in a region (as at the TeVatron). 

The other reason jet substructure has ballooned in recent years is because it is needed for many searches. For example, the high energy of the LHC means that particles such as top quarks are often produced not at rest, but with energies much larger than their mass. When such a highly boosted top quark decays into 3 jets (see Sec.~\ref{sec:top}), those jets will all be going in nearly the same direction. Thus a $t \bar{t}$ event at the LHC may look a lot like a dijet event~\cite{Kaplan:2008ie}. To distinguish the two event classes, a very powerful method is to start by finding {\bf fat jets}, of size $R\sim 1.0$ and then look for subjets of size $R\sim 0.4$ inside those jets~\cite{Butterworth:2008iy}. There are many ways by which this can be done, and new methods are continually being developed.

%
%
%

\section{Meet the Standard Model \label{sec:SM}}
In this section, we will go through the particles in the Standard Model, discussing how they decay and various things you should
know about them from a collider perspective. A good resource for a lot of this information is the PDG website
 \url{http://pdglive.lbl.gov/}.
 
\subsection{$W$  boson}
The $W$ boson has mass and width
\be
m_W = 80.385 \pm0.015~\GeV,\qquad \Gamma_W = 2.085\pm 0.042~\GeV
\ee
The $W$ boson mass is currently best measured through methods like fitting the $M_T$ distribution (see Section~\ref{sec:dist}). 


The $W$ decays through weak interactions with amplitudes proportional to the appropriate CKM elements. These elements are
\be
V_{\text{CKM}}=
\begin{pmatrix}
V_{ud} \sim 1 & V_{us} \sim 0.2 & V_{ub} \sim 0.003\\
V_{cd} \sim 0.2 & V_{cs} \sim 1 & V_{cb} \sim 0.04\\
V_{td} \sim 0.008 & V_{ts} \sim 0.04 & V_{tb} \sim 1
\end{pmatrix}
\label{ckm}
\ee
Decays of $W$ to top quarks are forbidden since the top quark is heavier than the $W$. Thus, there are only
two hadronic decay modes of the $W$ with order 1 rates. There are also the 3 leptonic decay modes.
\be
\underbrace{W^+ \to u \bar{d}}_{\times 3~\text{colors}},
\qquad \underbrace{ W^+\to c\bar{s}}_{\times 3~\text{colors}},
\qquad W^+ \to e^+ \nu,
\qquad W^+ \to \mu^+\nu,
\qquad W^+ \to \tau^+ \nu,
\ee
Because of the 3 colors of the hadronic decays, there are $3\times 2 + 3 = 9$ ``modes", of which $6/9  = 2/3$ are hadronic (jets) and 1/9 goes to each lepton flavor. This  rough estimate is consistent with the actual decay modes of the $W$:
\be
\hspace{2cm}
\begin{tikzpicture}
\node [above] at (-7,0) {\bf $\mathbf W$ boson};
\node  [below] at (-7,0) {\bf branching ratios}; 
  \pie[rotate=0, scale font ]{10/$e^\pm\nu$, 10/$\mu^\pm \nu$, 10/$\tau^\pm\nu$, 70/jets}
\end{tikzpicture}
\nonumber
\ee

\subsection{$Z$  boson}
The $Z$ boson has mass and width
\be
m_Z = 91.1876 \pm 0.0021~\GeV,\qquad \Gamma_Z =2.4952 \pm 0.0023~\GeV
\ee
Its couplings are flavor-diagonal and do not depend on the CKM elements. The $Z$ boson couples to fermions proportional to $T_3 - Q \sin^2\theta$, where $\sin^2\theta \sim 0.23$,  $T_3$ is the $SU(2)$ quantum number and $Q$
is the electric charge.  For the left and right handed fermions, the $Z$ boson couplings are
\be
\begin{tabular}{|ccc|c|}
\hline
LH particle & $T_3$ & $Q$ & $T_3 - Q \sin^2\theta$ \\
\hline
$\nu_e, \nu_\mu, \nu_\tau$  &$\frac{1}{2}$ & 0& 0.5\\
$e^-, \mu^-, \tau^-$  & $-\frac{1}{2}$ & -1& -0.28\\
$u,c,t$  & $\frac{1}{2}$ & $\frac{2}{3}$& 0.35 \\
$d,s,b$ & $- \frac{1}{2}$ & $-\frac{1}{3}$ &  -0.42 \\[1mm]
\hline
\end{tabular}\
\hspace{5mm}
\begin{tabular}{|ccc|c|}
\hline
RH particle & $T_3$ & $Q$ & $T_3 - Q \sin^2\theta$ \\
\hline
$\nu_e, \nu_\mu, \nu_\tau$  &$0$ & 0& 0\\
$e^-, \mu^-, \tau^-$  & $0$ & -1& 0.23\\
$u,c,t$  & $0$ & $\frac{2}{3}$& -0.15 \\
$d,s,b$ & $0$ & $-\frac{1}{3}$ &  0.08\\[1mm]
\hline
\end{tabular}
\ee
The $Z$ decay rate to a particular mode is proportional to $(T_3 - Q \sin^2 \theta_w)^2$.
Thus the branching ratio is given by  $(T_3 - Q \sin^2 \theta_w)^2$ for that mode divided
by the sum over $(T_3 - Q \sin^2 \theta_w)^2$ over all the modes (excluding top to which
the $Z$ cannot decay). This sum is $\Gamma_Z \propto 3.30$. Then the branching ratio
to electrons for example is
\be
\text{BR}(\Gamma \to e^+e^-) = \frac{(-0.28)^2 + (0.22)^2}{3.30}=3.5\%
\ee
to up quarks is
\be
\text{BR}(\Gamma \to u\bar{u}) =3 \times \frac{(0.35)^2 + (-0.15)^2}{3.30}=13\%
\ee
and so on.
The resulting decay modes are summarized in this chart:
%
%

\hspace{2cm}
\begin{tikzpicture}
\node [above] at (-7,0) {$\mathbf Z$ \bf boson};
\node  [below] at (-7,0) {\bf  branching ratios}; 
  \pie[rotate = 0,scale font]{3.3/$e^+ e^-$, 3.3/$\mu^+\mu^-$, 3.3/$\tau^+\tau^-$,  55/jets,  20/ $\nu \bar{\nu}$,15/$b\bar{b}$}
\end{tikzpicture}

Note that the $Z$ boson only decays around 7\% of the time to the ``golden" modes, $e^+e^-$ or $\mu^-\mu^-$. Most of the time it decay to jets.

\subsection{Top quark \label{sec:top}}
The pole mass and width of the top quark are
\be
m_t^\text{pole} = 173.1\pm 0.6~\GeV, \qquad \Gamma_t =1.41 \pm 0.19~\GeV
\ee

The top quark decays essentially 100\% of the time to a $b$ quark and a $W$. So its branching ratios
are determined by the $W$ branching ratios. 
We call a decay {\bf leptonic} if  the the $W$ decays to electrons or muons. Technically, $W\to \tau\nu$
is also leptonic, however tauons are not easy to see like electrons or muons, as they often decay hadronically (see Section~\ref{sec:tau}).
Generally, tops are produced in pairs through gluon fusion, $gg \to g \to t\bar{t}$ or from $q\bar{q} \to g \to t\bar{t}$. The $t\bar{t}$
system can they either decay leptonically, if both $W$'s decay to electrons or muons, {\bf semi-leptonically} if one decays to leptons and the other $W$ to hadrons, or {\bf hadronically} if both $W$'s decay to jets. The branching ratios are

\begin{tikzpicture}
\node [above] at (0,4) {\bf Top quark};
\node  [below] at (0,4) {\bf branching ratios}; 
  \pie[rotate = 150,scale font]{10/$b e^\pm \nu$, 10/$b \mu^\pm \nu$, 10/$b\tau^\pm \nu$, 70/jets}
\end{tikzpicture}
\begin{tikzpicture}
\node [above] at (-1,4) {\bf ${\mathbf t} \bar{ \mathbf{t} }$ };
\node  [below] at (-1,4) {\bf decay channels}; 
  \pie[rotate = 0,scale font]{40/semi-leptonic, 4/leptonic, 56/hadronic}
\end{tikzpicture}

You might think that the fully leptonic channel would be the best to study tops. However,
this channel has two drawbacks: its small rate  (4\% of the all $t\bar{t}$ pairs) and fact that with two leptonic decays, there are two neutrinos so the decay products cannot be completely reconstructed. Instead, the semi-leptonic channel proves the most useful. Nearly half the $t\bar{t}$ pairs decay semi-leptonically. More importantly however, the leptonic side can be fully reconstructed. The system triggers with
a hard electron or muon, missing energy and a $b$ jet. Then using the $W$ boson mass constraint, the neutrino momentum is determined. The three jets on the hadronic side can then be studied, for example, by using their combined invariant mass to find $m_t$. 

Measuring the top quark mass precisely is a formative challenge, with compelling theoretical and experimental issues. First of all, there is an ambiguity on what subtraction scheme the measured top quark mass corresponds to. If the mass is measured through something like the $t\bar{t}$ total or differential cross section, then one has good theoretical control over the scheme. In particular, the cross section can be calculated in $\msbar$ and extracted top mass is then the $\msbar$ mass.
Unfortunately, cross section measurements have large uncertainty, due to theoretical precision, statistics, backgrounds, and uncertainties on parton distribution functions (PDFs).

The top quark mass measurements with the smallest uncertainties come from fits to the invariant mass distribution of the hadronic top decay products in semi-leptonic $t\bar{t}$ events. Unlike cross section measurements, one does not need to know exactly how many tops are produced; instead one can put hard cuts and use the cleanest events. Moreover, precision knowledge of the PDFs is not necessary. These fits are done to simulations using Monte Carlo event generators. Thus the top mass measured is the Monte-Carlo mass, $m_t^{\text{MC}}$. This Monte Carlo mass is believed to be close to the top quark pole mass, although it has hard to make the correspondence exact. See~\cite{Moch:2014tta} for more details. 

%
%
\subsection{Bottom quark}
The bottom (or beauty) quark has an $\msbar$ mass of
\be
m_b(\mu=m_b) = 4.18 \pm 0.03~\GeV
\ee
Unlike the top quark, the bottom quark hadronizes before it decays. The hadrons are $B$-mesons and $B$-baryons. 
The most common $B$-hadrons are
\be
B^- = b\bar{u},\quad B^0 = b \bar{d}, \quad B_s = b \bar{s}
\ee
These $B$-hadrons decay through the weak decay of the bottom quark itself. The dominant decay mode of the bottom quark is
$b\to cW$. 

Let's try to estimate the $b$ lifetime using dimensional analysis. 
One way to do the estimate is to compare the $b$ decay to another weakly decaying particle, 
the muon, whose lifetime is around $2 \mu s$ (this is a number worth knowing, but also not hard to estimate by dimensional analysis). 
Weak decays are proportional to  $|\frac{g}{m_W^2}|^2 \sim G_F^2$. 
To get a rate, with mass dimension 1,
we must compensate by something with mass dimension 5. If the decay products' masses are negligible,
the only scale left is the decaying particles mass. So the rate scales like $m_\mu^5$ or $m_b^5$.
The muon can decay to $e^-\nu\nu$ only. However, the $b$ can decay to $c e^-\nu$ as well
as $c \mu^-\nu$, $c u \bar{d}$ and $c s \bar{c}$. Adding 3 colors for the hadronic channels, there
can be about 9 times as many decay modes from the $b$ as the $\mu$. 
We then have
\be
\Gamma_\mu \sim G_F^2 m_\mu^5,\qquad
\Gamma_b \sim 9 |V_{cb}|^2 G_F^2 m_b^5 
 =  9 |V_{cb}|^2\frac{m_b^5}{m_\mu^5} \times \Gamma_\mu
= 10^{-6}\times \Gamma_\mu
\label{bestimate}
\ee
The factor of 1 million means that the typical $B$-hadron lifetimes are in picoseconds $(10^{-12} s)$.  Thus,
the decay length (how far a $B$ hadron goes before decaying) is
\be
c\tau \sim 500 \times 10^{-6}~\text{m} \sim 0.5~\text{mm}
\ee
If the $b$ is produced relativistically, this decay length is longer due to time dilation. For example, an 50 GeV
$b$ from a top decay would have a Lorentz factor $\gamma \sim 10$ and its decay length might be
$\gamma c \tau \sim$ 5 mm.
Tagging of $b$ quark products at the LHC is predicated on being able to see these $\sim$mm decay lengths.

A typical decay chain of a $B$ meson is
\be
B^0 \to( D^+ \to (K^+ \to \pi^+ \pi^0)\pi^+\pi^-)\mu^- \nu
\ee
This final state has 4 charged particles. Having 4, 5 or 6 charged decay products is typical of a $B$ decay. Is it also
possible for $B$ mesons to decay to leptons ($\mu$ or $e$). This happens around 10\% of the time. There is
also a 10\% chance that a $D$ meson to which the $B$ decays will decay leptonically. Thus there is a 20\% chance of getting a lepton from a $B$ decay.

Typical $B$-tagging algorithms combine these observables:
\begin{enumerate}
\item Look for a relatively high multiplicity of tracks.
\item Look for a displaced (secondary) vertex. That is, look for tracks which converge on  a point separated by
a distance $d\sim 0.5$ mm from the primary vertex.
\item If the secondary vertex cannot be resolved, one can instead use the impact parameters of the various tracks
which do not converge on the primary vertex. The impact parameter is the shortest distance between the
line represented by a track and the primary vertex. 
\item Construct the invariant mass of the particles converging on the secondary vertex. Look for this to be
close to $m_b$. 
\item Combine the relevant observables with simple cuts or more typically using a neural network or boosted decision tree. 
\end{enumerate}

For some rough efficiency numbers, current $b$ tagging algorithms can keep 70\% of the $b$'s while rejecting 
light quark jet ($u, d, s$-initiated) backgrounds by a factor of 50. Generally, gluon jets and charm quark jets
are harder to distinguish from $b$ jets than light quark jets. 

Bottom quark mesons have a very heavy $b$ quark surrounded by a light quark. This system can be studied in good
analogy with the hydrogen atom -- the heavy proton is surrounded by a light electron. For example, relations
between ground state $B$ mesons and angular excitations, such as the $B^\star$ (spin 1) can be derived
up to corrections in $\LQCD/m_b$. There is a very powerful effective field theory for studying
$B$ hadrons called Heavy Quark Effective Theory (HQET). 

Bottom quarks also form $b\bar{b}$ bound states called {\bf bottomonium}. The lightest is the 
$\Upsilon(1S)$ 
with a mass of $m_{\Upsilon(1S)} = 9460$. The $\Upsilon(1S)$ is too light to decay to $B$ mesons. There are other Upsilon particles,
analogous to the excited states for the hydrogen atom. The $\Upsilon(4S)$ has mass $m_{\Upsilon(4S)} = 10.5$ GeV. It
almost exclusively decays to $B\bar{B}$. Thus the B-factories (Belle and BaBar) would run at the center-of-mass energy
to resonantly produce the $\Upsilon(4S)$. This procedure made an endless supply of $B$ mesons for precision study. 

\subsection{Charm quark}
The charm quark has an $\msbar$ mass of
\be
m_c(\mu=m_c) = 1.29 \pm 0.1~\GeV
\ee
Charm mesons are called ``D" mesons (these are not mesons with a down quark in them!). For example,
\be
D^+ = c\bar{d},\quad D^0 = c \bar{u}, \quad D_s = c \bar{s}
\ee
A similar estimate to Eq.~\eqref{bestimate}, using $|V_{cs}| \sim 1$  gives
\be
\Gamma_c \sim 
   5 \frac{m_c^5}{m_\mu^5} \times \Gamma_\mu
= 10^{-6}\times \Gamma_\mu
\label{cestimate}
\ee
So we also get picosecond lifetimes for charm mesons.  It turns out order-one numbers are important
here and the typical decay length for a $D$ meson is around 300 $\mu$m. Thus charmed mesons
travel about 60\% as far as bottom mesons before they decay. 

Charm-jet tagging is generally harder than bottom-jet tagging for various reasons:
\begin{itemize}
\item The decay length of charm is shorter than bottom, so it is harder to resolve the displaced vertex and
the impact parameters of the displaced tracks are smaller.
\item Charm hadron decays have lower multiplicity than bottom hadron decays -- fewer charged particles means less information to 
work with in charm tagging.
\item $B$ mesons almost always decay to $D$ mesons, so there is usually a charm in a bottom decay.
 Thus distinguishing charm from bottom presents extra challenges. 
\end{itemize}

On the other hand, even if we could tag charm, it would not be terribly useful for Higgs or top physics at the LHC. In contrast to $b$-jets,
which are important for finding top quarks  and the dominant $H\to b\bar{b}$ decay mode,
charm jets are not dominant in any Standard Model process. Eventually, it would be nice to measure the $H\to c\bar{c}$
branching ratio, but this will be very challenging at the LHC and is probably impossible without an inverse attobarn of data at minimum. Charm quarks are mostly of interest for precision flavor physics that can provide indirect evidence of new particles. 

One charm meson of particular interest is the $J/\Psi$ particle. This particle has two names because it was discovered
independently by two groups within a few months of each other. The $J/\Psi$ is the lightest {\bf charmonium} state, that is,
it is a $c\bar{c}$ bound state. It has a mass of 3 GeV and an extremely narrow width of 93 keV. It decays about 5\% of the 
time to $e^+ e^-$ and 5\% to $\mu^+ \mu^-$. Because of its narrow width and purely leptonic decays it plays a special
role as a standard candle at colliders, useful for energy calibration for example.

\subsection{Strange quark}
The strange quark has an $\msbar$ mass of
\be
m_s(\mu=m_s) = 95 \pm 5~\MeV
\ee
Strange mesons are Kaons:
\be
K^- = s\bar{u},\quad K^0 = s \bar{d},\quad \bar{K}^0 = d\bar{s}, \quad K^+ =  \bar{s} u
\ee
The neutral kaons, $K^0$ and $\bar{K}^0$ have the same quantum numbers and can mix. 
They can decay to 2 pions or 3 pions. The 2 pion state is CP even and the 3 pion state is CP odd. Since CP is conserved in QCD, it is natural to use CP eigenstates
\be
K_1 = \frac{K^0 + \bar{K}^0}{\sqrt{2}}\to \pi^+ \pi^- ~~(\text{CP even})
\qquad
K_2 = \frac{K^0 - \bar{K}^0}{\sqrt{2}}\to \pi^+ \pi^- \pi^0 ~~(\text{CP odd})
\ee
Because CP is violated by the weak interactions, $K_2$ can sometimes decay to $\pi^+\pi^-$ determined by
a parameter $\epsilon^\prime\sim10^{-6}$. In addition, CP violation
implies that  $K_1$ and $K_2$ are not  exact mass eigenstates. The mass eigenstates are
\be
K_L = K_2 - \epsilon K_1,\qquad
K_S = K_1 + \epsilon K_2
\ee
These are the physical particles, called ``$K$-long'' and ``$K$-short''. At the LHC, we are not really sensitive
to the CP violation. So essentially there are two neutral kaons, $K_L$ with a lifetime  of $\tau\sim10^{-8} s$,
similar to the $K^\pm$ lifetimes, and a decay length of $\gamma c\tau_L \gtrsim 3$ m (depending on boost) and $K_S$ with a lifetime
of $10^{-10}s$ and a decay length of $\gamma c\tau_S \gtrsim 3$ cm.

\subsection{Up and down quarks}
Up and down quarks are effectively massless at the LHC. Their bound states are pions
\be
\pi^+ = u\bar{d},\quad \pi^0 = \frac{1}{\sqrt{2}}(u\bar{u} - d\bar{d}),\qquad \pi^- = d\bar{u}
\ee
The charged pions have lifetimes of $\tau \sim 10^{-8} s$ and decay lengths of $\gamma c\tau_\pi   \gtrsim  8$ m.
There are also baryonic bound states of up and down quarks, such as the proton and neutron. 

Charged pions decay to muons and neutrinos. However, they are considered stable on the time and length scales of the experiment. 
Charged pions leave charged tracks  and deposit some of their energy in the ecal and most of
their energy in the hcal. Neutral pions decay to photons $\pi^0 \to \gamma \gamma$. This decay occurs through a quark
triangle diagram and is a strong, not a weak decay. The lifetime of $\pi^0$ is therefore much shorter than that of the $\pi^\pm$, around $10^{-17}$ s. Thus the $\pi^0$ particles are not considered stable; they decay before
they reach the detector and show up as two photons in the ecal.

%

\begin{table}
{\small
$
\begin{array}{ccccccccccccc}
 \text{$<$event$>$} & \text{} & \text{} & \text{} & \text{} & \text{} &
   \text{} & \text{} & \text{} & \text{} & \text{} & \text{} & \text{} \\
\text{\# particle code} &~~ \text{stable?} &\multicolumn{2}{c}{\text{mothers}} & \multicolumn{2}{c}{\text{colors}}  &  p_x & p_y & p_z & E & m\\
 21 & -1 & 5 & 0 & 504 & 501 & 0. & 0. & 2572.17 & 2572.17 & 0.0 \\
 21 & -1 & 6 & 6 & 503 & 502 & 0. & 0. & -878.377 & 878.377 & 0.0 \\
 21 & 2 & 1 & 2 & 503 & 501 & -739.1 & -1073.06 & 1707.08 & 2147.52 & 0.0 \\
 21 & 2 & 1 & 2 & 504 & 502 & 739.1 & 1073.06 & -13.28 & 1303.03 & 0.0 \\
 \multicolumn{3}{l}{\cdots~\text{unstable particles}~\cdots}
 \\
  111 & 1 & 1314 & 1315 & 0 & 0 & -0.681104 & 0.244785 & 7.92263 & 7.95676 & 0.13498 \\
 211 & 1 & 1318 & 1319 & 0 & 0 & -0.167889 & 0.717687 & 8.73263 & 8.76479 & 0.13957 \\
 -211 & 1 & 1322 & 1323 & 0 & 0 & 2.04007 & 1.73955 & -363.582 & 363.592 & 0.13957 \\
 211 & 1 & 1337 & 1340 & 0 & 0 & 0.0266408 & -0.227236 & -40.8654 & 40.8662 & 0.13957 \\
 211 & 1 & 1325 & 1332 & 0 & 0 & -0.541204 & 1.10963 & 636.361 & 636.362 & 0.13957 \\
 111 & 1 & 1325 & 1332 & 0 & 0 & -0.197198 & 0.67533 & 338.412 & 338.413 & 0.13498 \\
 2112 & 1 & 1325 & 1332 & 0 & 0 & 0.0309196 & -0.662625 & 1007.1 & 1007.1 & 0.93957 \\
 -321 & 1 & 1353 & 1366 & 0 & 0 & -1.25276 & -0.859944 & -1.18997 & 1.99215 & 0.49368 \\
 211 & 1 & 1353 & 1366 & 0 & 0 & 0.273685 & -0.308268 & -0.0785492 & 0.442247 & 0.13957 \\
 -321 & 1 & 1353 & 1366 & 0 & 0 & 2.13548 & 3.85915 & 0.121846 & 4.43981 & 0.49368 \\
 111 & 1 & 1353 & 1366 & 0 & 0 & 0.0304648 & 0.328838 & 0.327497 & 0.48429 & 0.13498 \\
 -321 & 1 & 1353 & 1366 & 0 & 0 & -0.595369 & 0.622589 & 0.41675 & 1.07679 & 0.49368 \\
 211 & 1 & 1353 & 1366 & 0 & 0 & -0.265771 & -0.115541 & -0.0389015 & 0.324001 & 0.13957 \\
 211 & 1 & 1383 & 1394 & 0 & 0 & 0.262096 & 0.17178 & -20.3439 & 20.3468 & 0.13957 \\
 \multicolumn{3}{l}{\cdots~\text{400 more particles}~\cdots} \\
111 & 1 & 1893 & 0 & 0 & 0 & 0.0211716 & -0.0976666 & -0.578813 & 0.602686
   & 0.13498 \\
 -2112 & 1 & 1925 & 0 & 0 & 0 & 1.41653 & -0.27262 & -32.9999 & 33.0448 &
   0.93957\\
 111 & 1 & 1925 & 0 & 0 & 0 & 0.413614 & 0.0268691 & -8.37944 & 8.39077 &
   0.13498 \\
 2212 & 1 & 1942 & 0 & 0 & 0 & -0.912913 & 0.0206735 & 3.18503 & 3.44363 &
   0.93827 \\
    \text{$<$/event$>$} & \text{} & \text{} & \text{} & \text{} & \text{} &
   \text{} & \text{} & \text{} & \text{} & \text{} & \text{} & \text{} \\
\end{array}
$
\begin{center}
\begin{tabular}{|ccc|}
\hline
particle & code & \# in event \\
$\pi^\pm$ & $\pm 211 $ & 205 \\
$\pi^0$ & 111 & 124 \\
$n$ & 2112 & 25 \\
$K^\pm$ & $\pm 321$ & 18 \\
\hline
\end{tabular}
\begin{tabular}{|ccc|}
\hline
particle & code & \# in event \\
$\gamma$ & $22$ & 14  \\
$p^\pm$ & $\pm 2212$ & 11 \\
$K_L$ & 130 & 4 \\
$e^\pm$ & $\pm11$ &  2 \\
\hline
\end{tabular}
\end{center}
}
\caption{Example $m_{jj} =  3$ TeV dijet event at the LHC. The first 4 lines with numbers are the $gg\to gg$ partonic process (21 is a gluon). 
Common particle codes are shown in the box. As you can see, most of the particles in this, and
most events, are pions. } \label{tab:event}
\end{table}

An example list of particles produce in  an LHC event is shown in Table~\ref{tab:event}. This is the output from {\sc pythia} for a dijet process, $gg\to gg$
at the parton level, with $\hat{s} = 3$ TeV.  This particular event has 403 particles, of which 329, or 80\%, are pions. 
The $\pi^0$'s decay promptly to photons, but they have not been decayed in this table for
pedagogical reasons. For example, you can  see that there are about twice as many
charged pions as neutral ones. This follows simply from isospin, or more simply from there being two charged pions
but only one neutral one. Thus about 2/3 of particles produced at the LHC leave charged tracks. 

\subsection{Electrons and muons}
I don't have too  much to say about electrons and muons. Electrons are light so they radiate a lot. They leave tracks and deposit all their energy in the electromagnetic calorimeter.

Muons have  mass $m_\mu = 105$ MeV. The muon lifetime is $\tau = 2\times 10^{-6}$ s which gives a decay length $\gamma c \tau \gtrsim 300$ m.  Thus muons are stable as far as the experiment is concerned. Because muons leave the detector before depositing all their energy, their momenta must be measured from the curvature of their trajectories in a magnetic field. That's why the muon detectors and magnetic fields at ATLAS and CMS are so big -- hard muons have very straight tracks, so we need these strong fields
and large detectors to see enough curvature to measure the muons' momenta. 

\subsection{Tauons \label{sec:tau}}
The $\tau$ lepton has a mass $m_\tau = 1.77$ GeV. It decays through the weak force. 
Rescaling the factors from the muon decay rate, as in Eq.~\eqref{bestimate}, gives basically the same thing as
for charm, Eq.~\eqref{cestimate}, 
\be
\Gamma_\tau \sim 
   5 \frac{m_\tau^5}{m_\mu^5} \times \Gamma_\mu
= 10^{-7}\times \Gamma_\mu
\ee
As for charm, the factor of $5$ comes from one hadronic channel with 3 colors and 2 leptonic channels. 
So the $\tau$ decays around $\sim 20\%$ of the time to electrons, $\sim 20\%$ to muons and $\sim 60\%$ to
hadrons. 
This rate implies $\tau_\tau \sim 10^{-13} s$, so $\tau$ decays are prompt. 

In leptonic $\tau$ decays, the final state is $\mu\nu \bar{\nu}$ or $e \nu \bar{\nu}$. Since the $\tau$ decays promptly (before
it leaves a track), these final states are essentially indistinguishable from promptly-produced electrons or muons. Sometimes  leptonic decays are still the best indication of a tauon -- for example, in $h\to\tau\tau$, the leptonic
modes are cleaner than the hadronic ones and $h\to e^+e^-$ and $h \to \mu^+\mu^-$ have negligible rates. But a general
$\tau$-tagger, which hopes to distinguish tauons from electrons and muons as well, must utilize the hadronic channels.

We estimated a  60\% branching ratio to hadrons. The actual rate is closer to  65\%.
Breaking down the 65\%, around 14\% goes to final states with 3 charged particles and around 51\% goes to final states with one charged particle. The number of charged particles implies a number of tracks or {\bf prongs} in the decay:

\hspace{2cm}
\begin{tikzpicture}
\node [above] at (-7,0) {\bf $\mathbf \tau$ lepton};
\node  [below] at (-7,0) {\bf branching ratios}; 
  \pie[rotate=0, scale font ]{51/ 1 prong, 14/3 prong, 35/leptonic}
\end{tikzpicture}

\noindent An example of a 3-prong decay is
\be
\tau^+ \to\nu (a_1^+ \to \pi^+\pi^-\pi^+) = \nu \pi^+\pi^-\pi^+  \qquad\text{ 3-prong decay}
\ee
Examples of the 1-prong decays 
are
\be
\tau^+ \to \pi^+ \nu,\qquad \tau^+ \to\nu (\rho^+\to \pi^+ ( \pi^0\to\gamma\gamma))= \nu \pi^+\gamma\gamma
 \qquad\text{ 1-prong decays}
\ee

The hadronic decays of tauons produce tiny little jets, with only a handful of particles in them. The procedure for finding tauons, $\tau$-tagging, involves looking for
\begin{itemize}
\item Low multiplicity ``jets" -- 1 or 3 prongs
\item Narrow jets
\item Isolated jets (not many hadrons nearby)
\end{itemize}
This last point is important -- because tauons are leptons, they are not generally produced in the context of lots of other QCD radiation, for example, from within a gluon jet. Thus one generally does not expect much hadronic radiation in the vicinity of the $\tau$. 

\subsection{Higgs boson \label{sec:higgs}}
Finally we come to the Higgs boson. The Higgs boson has a mass
\be
m_h = 125.09\pm 0.3 ~\GeV
\ee
Note that the uncertainty on the Higgs boson mass is at the sub-percent level already.

The Higgs can be produced in various ways. The cross sections for the various production 
channels at 13 TeV are
\begin{itemize}
\item Gluon fusion $\sigma(gg\to  h) \sim 44$ pb. 
\item Vector boson fusion $\sigma(pp \to qq h) \sim 4$ pb.
\item Associated production: $\sigma(pp \to W h) \sim 1.5$ pb. 
\item Associated production: $\sigma(pp \to Z h) \sim 0.88$ pb. 
\item Associated production: $\sigma(pp \to t\bar{t} h) \sim 0.5$ pb. 
\item Associated production: $\sigma(pp \to b\bar{b} h) \sim 0.5$ pb. 
\end{itemize}
At 8 TeV, the cross sections are about half of these numbers. For example, in run 1, the LHC accumulated 25 fb${}^{-1}$ of
data which amounts to 20 pb$\times 25~\text{fb}^{-1} = 500,000$ Higgs bosons produced. 

Most of these production channels and rates are straightforward to compute.
Since there is no gluon-gluon-Higgs interaction in the Standard Model, gluon fusion proceeds
through a top loop triangle diagram, as in Eq.~\eqref{ggh}. This diagram can be computed exactly, but it is often computed to leading order in $\frac{1}{m_t}$ (sometimes  called the $m_t \to \infty$ limit). In this limit, the production rate is equivalent to what one would get from the operator
\be
\begin{gathered}
\resizebox{30mm}{!}{
     \fmfframe(0,0)(0,0){
\begin{fmfgraph*}(40,30)
\fmfstraight
	\fmfleft{L1,L2}
	\fmfright{R}
	\fmf{gluon}{L1,v1}
	\fmf{gluon}{L2,v2}
	\fmf{fermion,tension=0.5}{v1,v2,v3,v1}
	\fmf{phantom,tension=0,label=$t$}{v2,v3}
	\fmf{dashes,label=$h$,l.s=left}{v3,R}
\end{fmfgraph*}
}}
\end{gathered}
\longrightarrow~~ \frac{\alpha_s}{12\pi}\frac{h}{v} G_{\mu\nu}^a G^a_{\mu\nu} 
\ee
Note that there is no dependence on the top mass at leading order in $\frac{1}{m_t}$ -- the net factor of $\frac{1}{m_t}$ from the top propagators (or dimensional analysis) multiplies the top Yukawa on the Higgs vertex leaving only the Higgs vev $v= \frac{m_t}{\lambda_t}$. The $m_t\to \infty$ limit is a very good approximation.  

The vector boson fusion channel is worth some discussion. In this channel, the Higgs boson is produced through the
$t$-channel exchange of two $W$ bosons which fuse to form the Higgs. 
\be
\begin{gathered}
\resizebox{30mm}{!}{
     \fmfframe(0,0)(0,0){
\begin{fmfgraph*}(40,30)
\fmfstraight
	\fmfleft{L1,L2}
	\fmfright{R1,R2,R4}
	\fmf{fermion}{L1,v1,R1}
	\fmf{fermion}{L2,v2,R4}
	\fmf{photon,label=$W^-$,l.d=0.1}{v1,v}
	\fmf{photon,label=$W^+$,l.d=0.1}{v2,v}
	\fmf{dashes,label=$h$}{v,R2}
\end{fmfgraph*}
}}
\end{gathered}
\ee
Because the process is $t$-channel, the cross
section is dominated by the kinematic region where $t$ is small, which is where the outgoing quarks remain close to the beam.
Thus a signature of vector boson fusion is two forward jets. Because vector boson fusion involves the scattering of vector
bosons, it is also useful for looking for anomalous triple gauge boson couplings.

The Higgs boson decays around 60\% of the time to $b\bar{b}$. Its branching ratios are

\hspace{2cm}
\begin{tikzpicture}
\node [above] at (-7,0) {\bf Higgs boson};
\node  [below] at (-7,0) {\bf branching ratios}; 
  \pie[rotate=0, scale font,sum=auto, after number = \% ]{60/$b\bar{b}$, 21/$W W^\star$, 9/$gg$, 6/$\tau\tau$,3/$Z Z^\star$}
\node [right] at (3,0) {0.1\% $\gamma\gamma$};
\end{tikzpicture}

\begin{figure}[t]
\begin{center}
  \includegraphics[width=0.33\textwidth]{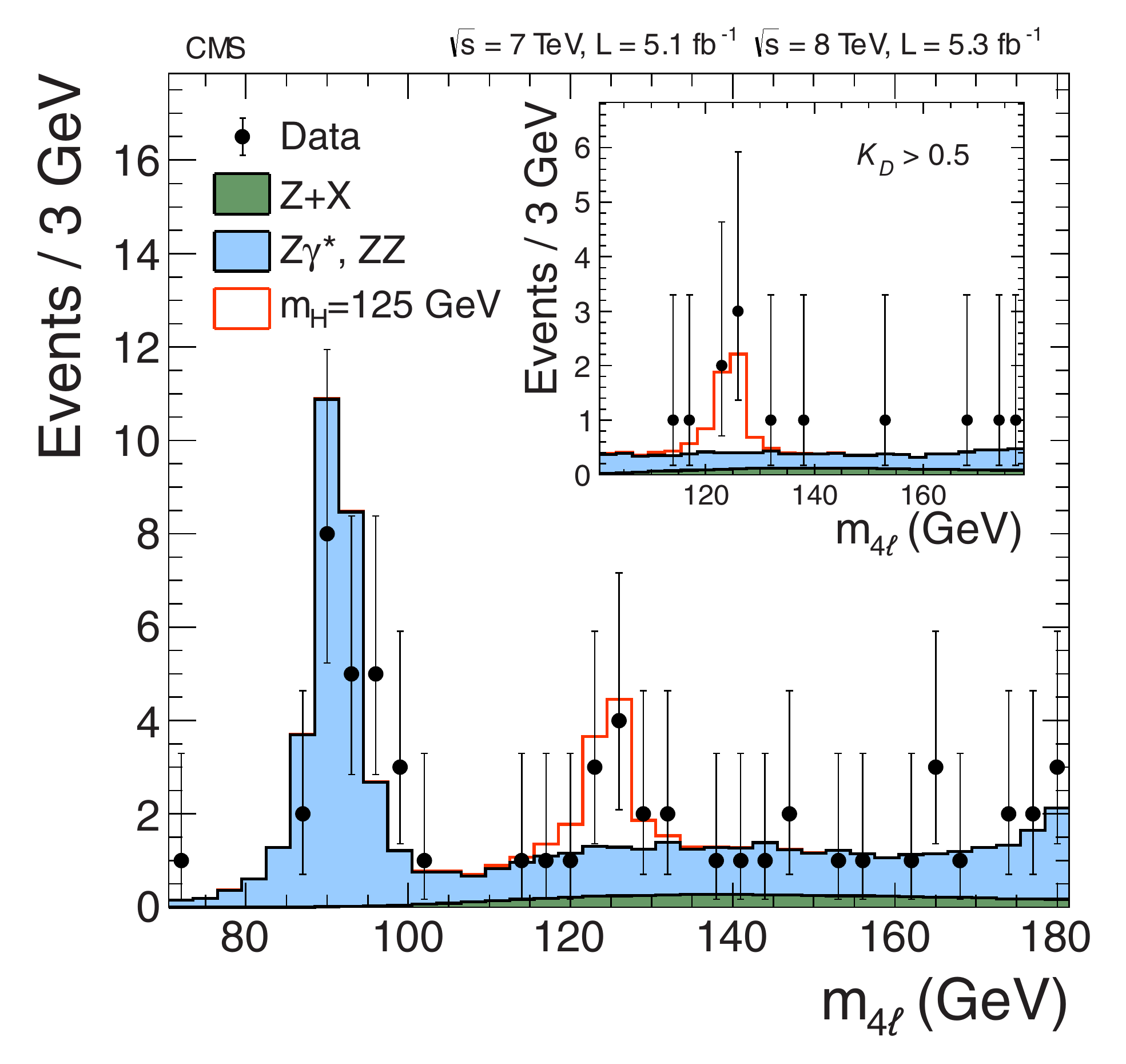}
 \includegraphics[width=0.32\textwidth]{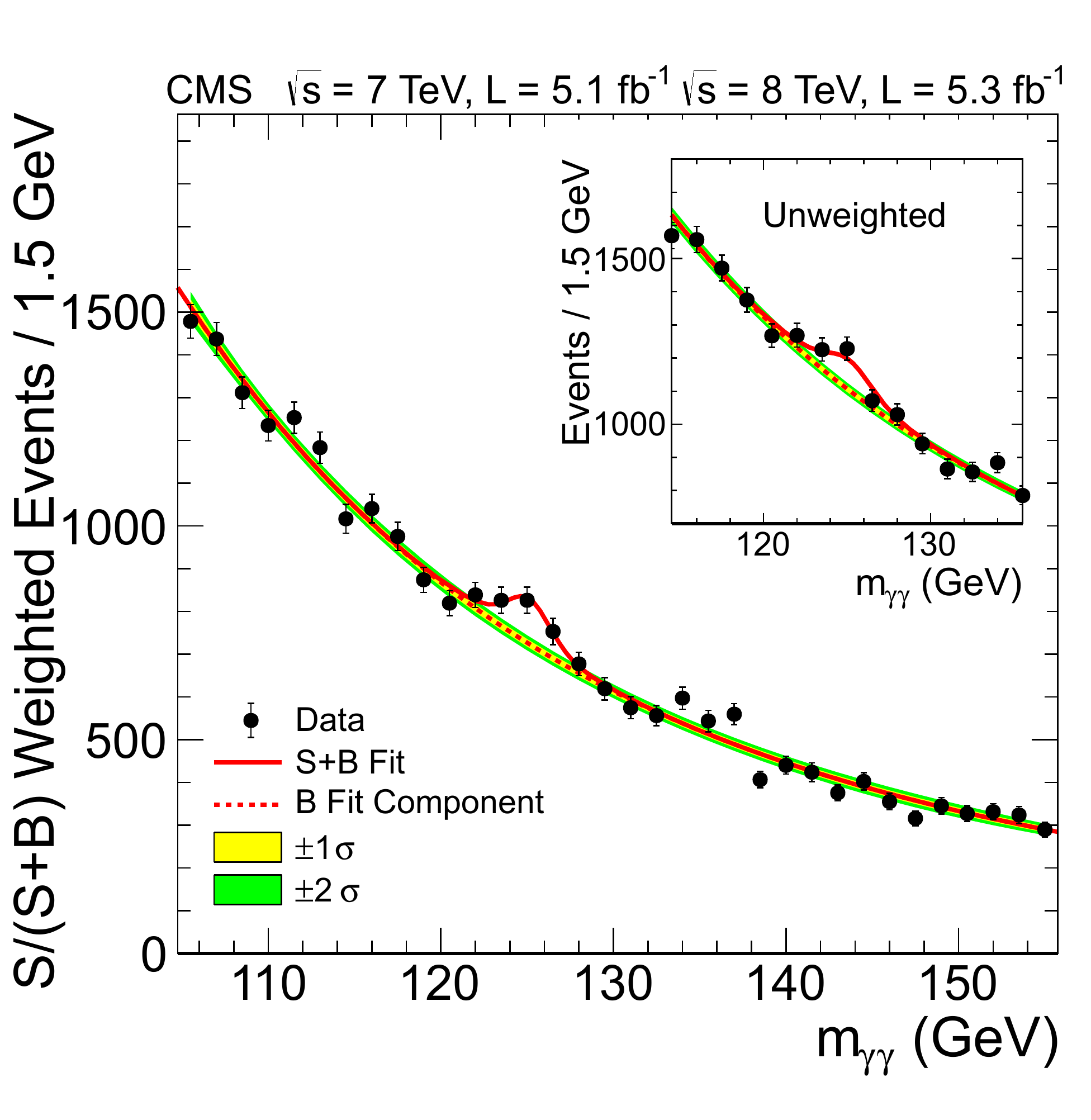}  
 \includegraphics[width=0.33\textwidth]{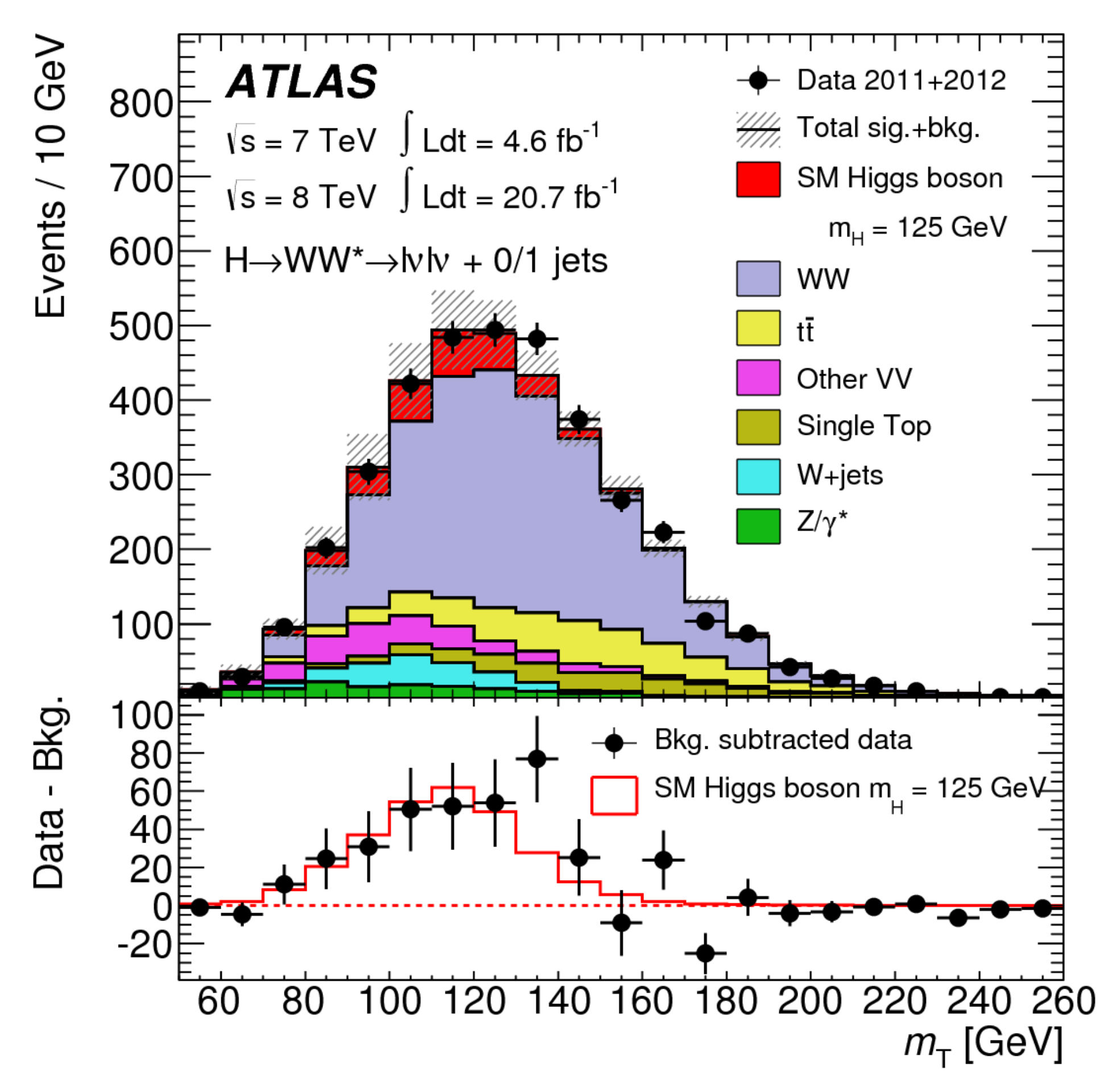}  
  \caption{
Higgs boson observations in various decay modes. The left two plots are discovery plots from
CMS~\cite{Altheimer:2012mn}:
 left shows the 4 lepton ``golden channel" and middle shows the digamma channel. The right plot is a transverse mass distribution from ATLAS showing the $h\to 2l2\nu$ decay mode~\cite{Aad:2012tfa}. 
  }
  \label{fig:higgs} 
\end{center}
\end{figure}

The ``golden channel" for Higgs is $h\to Z Z^\star \to 4l$, where $l=\mu$ or $e$ (the $\star$ on one of the $Z$'s means it is off-shell,
as it must be since $m_h < 2 m_Z$). This channel is golden not only 
 because there little background and the Higgs boson can be fully reconstructed, but also because the angular dependence of the 4
leptons can be used to determine the spin of the Higgs. 
Unfortunately, as $h\to Z Z^\star$ happens in only
3\% of Higgs decays, and as the $Z$ decays leptonically only 6\% of the time, the $h\to4l$ branching ratio is a measly $10^{-4}$.  
Thus, out of the 500,000 Higgs at run 1 there were about 50 $h\to 4l$ events. Accounting for experimental efficiencies, this number reduces further to the handful of events seen, 
as shown in Fig.~\ref{fig:higgs}.

The next cleanest channel is $gg \to h \to \gamma\gamma$. Here, the branching ratio is around 0.1\%, so we get 500 Higgs bosons decaying this way. Unlike $h\to 4l$, this channel has substantial background from $pp\to \gamma\gamma$ in QCD. 
The first Higgs discoveries in this channel were based on looking for a bump on top of a smooth background, independent of what the theory prediction was for that background. That's often how  first discoveries are made, since it's satisfying to see a bump by eye. The technique works if you can guarantee that the observable is smooth outside of the signal region,
which is hard to do without any theory input. Looking at the middle panel of Fig.~\ref{fig:higgs}, you clearly see a bump.
In fact, CMS has taken some editorial liberties to make this bump more apparent. First, they weight the data by their expectation for $S/(S+B)$. This enhances the signal region. Second, they add some lines
to guide the eye. Imagine the $h\to\gamma\gamma$ plot in Fig.~\ref{fig:higgs} without the green and red lines -- would you be able to definitively locate the Higgs mass peak? Contrast this with the $h\to4l$ signal where the peak is unmistakable.

Next we turn to the third Higgs boson discovery mode, $h\to W W^\star \to 2l2\nu$. This $h\to W W^\star$ decay
has a 40\% branching ratio, then demanding $e\nu$ or $\mu\nu$ final states has a $(20\%)^2$ branching ratio, giving 1.6\%
of Higgs bosons decaying this way. So of the 500,000 Higgs bosons produced, we get 800 $h\to 2l\nu\nu$ decays. Unfortunately,
the Higgs boson cannot be fully reconstructed this way, and so we cannot just look for a bump over background. Instead,
variables like transverse mass are used. An ATLAS measurement of $m_T$ is Higgs candidate events is shown in Fig.~\ref{fig:higgs}. Importantly, it is only possible to see a Higgs in this channel if
the backgrounds are known, including accurate theoretical predictions for their cross sections.

In summary, the are three Higgs` `discovery" channels. First, the golden channel $h\to Z Z^\star \to 4l$ which had around 50 events in run 1.
In this golden channel, $S/B$ is large and one can fully reconstruct the Higgs,  allowing its spin to be measured.
Next, $h\to\gamma\gamma$ had $\sim 500$ events, small $S/B$, but the Higgs could be fully reconstructed producing
a visible bump in $m_{\gamma\gamma}$ with enough statistics. This bump could be seen over a smooth fit to the sideband
region, independent of theory predictions.
 Then, $h\to WW^\star \to 2l2\nu$ had 5000 events, but one could not see the Higgs as a bump.
Instead there is a broad excess, so the Higgs can only be seen if the backgrounds are known.

All the above channels have the Higgs decaying to vector bosons. The fermionic decay modes of the Higgs are harder to see. 
It is nearly impossible to see $h\to b\bar{b}$ if the Higgs boson is produced from gluon fusion. The problem is that there is a
QCD $pp \to b\bar{b}$ background whose cross section is many orders of magnitude larger. The best place to see $h\to b\bar{b}$ is in the associated production channels, $pp \to Zh$, $pp \to W h$ and $pp\to t\bar{t}h$, or in vector boson fusion,
where the forward jets help reduce backgrounds. $h\to\tau\tau$ can also, in principle, be seen in these channels. In particular,
with vector boson fusion, since $h\to \tau\tau$  is an electroweak decay, one expects a relative absence of QCD radiation
in the central region compared to QCD backgrounds, 2 forward jets, and two low-multiplicity central jets from the
tauons. 

\section*{Acknowledgements} 
I would like to thank the organizers of TASI, Rouven Essig, Ian Low and  Tom DeGrand for inviting me to lecture, and to the students for all
their excellent questions and discussions. I would also like to thank the Galileo Galilei Institute for inviting me as well, where a version of these lectures were also given.
My work is conducted with support in part from the U.S. Department of Energy under contract DE-SC0013607. 

\bibliography{tasi}

\bibliographystyle{utphys}

\end{fmffile}
\end{document}